\documentclass[aip,jcp,reprint,eqsecnum,superscriptaddress,twocolumn,longbibliography]{revtex4-1}

\usepackage{natbib}
\usepackage{amssymb,amsmath}

\usepackage{epsfig}
\usepackage{bm}
\usepackage{color}
\usepackage{graphicx}% Include figure files
\usepackage{hyperref}
\usepackage{enumerate}

\newcommand\beq{\begin{equation}}
\newcommand\eeq{\end{equation}}
\newcommand\beqa{\begin{eqnarray}}
\newcommand\eeqa{\end{eqnarray}}
\newcommand{\nn}{\nonumber\\}
\def\bal#1\eal{\begin{align}#1\end{align}}
\newcommand{\thr}{{\text{th}}}
\newcommand{\Mp}{{\text{Mp}}}
\newcommand{\mbf}{m_{\text{bf}}}
\newcommand{\sbf}{\sigma_{\text{bf}}}
\newcommand{\nbf}{n_{\text{bf}}}
\newcommand{\ancho}{0.75\columnwidth}

\newcommand{\st}{\text{s}}

\begin{document}

\title{Mpemba effect in molecular gases under nonlinear drag}

\author{Andr\'es Santos}
\email{andres@unex.es}

\affiliation{Departamento de F\'{\i}sica and Instituto de
  Computaci\'on Cient\'{\i}fica Avanzada (ICCAEx), Universidad de
  Extremadura, 06006 Badajoz, Spain}

\author{Antonio Prados}
\email{prados@us.es}

\affiliation{F\'{\i}sica Te\'orica, Universidad de Sevilla, Apartado de Correos 1065, 41080
Sevilla, Spain}

\date{\today}

\begin{abstract}

We look into the Mpemba effect---the initially hotter sample cools sooner---in a molecular gas with nonlinear viscous  drag. Specifically, the gas particles interact among them via elastic collisions and also with a background fluid at equilibrium. Thus, within the framework of kinetic theory, our gas is described by an Enskog--Fokker--Planck equation. The analysis  is carried out in the first Sonine approximation, in which the evolution of the temperature is coupled to that of excess kurtosis. This coupling leads to the emergence of the Mpemba effect, which is observed in an early stage of relaxation and when the initial temperatures of the two samples are close enough. This allows for the development of a simple theory, linearizing the temperature evolution around a reference temperature---namely the initial temperature closer to the asymptotic  equilibrium value. The linear theory provides a semiquantitative description of the effect, including expressions for  crossover time and  maximum temperature difference. We also discuss the limitations of our linearized theory.

\end{abstract}

\maketitle

\section{Introduction}
\label{sec1}

One of the signatures of nonequilibrium systems is the presence of
memory effects.\cite{KPZSN19} A system displays memory when its time
evolution from a given initial state is not uniquely determined by the
initial values of its macroscopic---or hydrodynamic variables. In
other words, the system evolution depends on how it has been
previously being aged; memory effects are thus intimately related to
aging,\cite{B92} which has been typically associated with glassy
behavior.\cite{BP93,BP94,ANMMM00,RS03,JMP19}
Notwithstanding this, in addition to being investigated in models for
glasses,\cite{KH89,CKR94,PBSR97,BB02}
it has been found in many different physical systems: granular
fluids,\cite{BPGM07,AP07} dense granular
matter,\cite{JTMJ00,BP01b}
ferroelectrics,\cite{HCCW17} disordered mechanical
systems,\cite{LGAR17} and frictional interfaces,\cite{DS18} to name
just a few.

The Mpemba
effect\cite{MO69,LR17,LVPS17,KRHV19,Betal19,YH20} is a
counterintuitive memory phenomenon: given two samples of fluid, the
one that is initially hotter may cool more rapidly. Therefore, the
curves describing the time evolution of the temperature for the two
samples cross each other at a certain time $t_{c}$, and the curve for
the initially hotter sample stays below the other one for longer
times, $t>t_{c}$.  It is important to characterize the range of values
of the relevant physical quantities that allow for the emergence of
the Mpemba effect; in general, the difference of initial temperatures
must be small enough.

Although first reported in the case of
water,\cite{aristotle_works_1931,MO69} its existence for that liquid
is still controversial.\cite{BL16,GLH19} As a proof of concept,
the feasibility of the Mpemba effect has recently been reported in
granular gases.\cite{LVPS17,TLLVPS19,BPRR20,MLTVL20} Therein, collisional
inelasticity couples the evolution of the (granular) temperature to
other quantities---such as the kurtosis or the
rotational-to-translational energy ratio---monitoring the
nonequilibrium nature of the velocity distribution function (VDF), even in
homogeneous and isotropic states.\cite{VS15,LVGS19}

In this work, we show that the Mpemba effect is also present in
homogeneous and isotropic states of molecular gases---i.e., with
elastic collisions---driven by an external drag force with a
velocity-dependent friction coefficient. The particles of our system  are supposed to be hard spheres, for the sake of simplicity,
surrounded by a background fluid in equilibrium. Gas particles collide
among them, these collisions being modeled by a Boltzmann--Enskog
collision term in the evolution equation for the VDF.

Gas particles also interact with the background fluid. This
interaction translates into two forces: (i) a macroscopic,
deterministic, nonlinear drag force and (ii) a stochastic force. The
intensity of the latter follows from the fluctuation--dissipation
theorem, which ensures that the gas VDF tends to a Maxwellian with the
temperature of the background fluid in the long-time limit. In the
evolution equation, the interaction between the gas and the fluid is
described by a Fokker--Planck term; therefore, the VDF  obeys an
Enskog--Fokker--Planck equation with nonlinear drag.

The framework of our work is thus nonlinear Brownian motion,\cite{K94}
but the Brownian particles are no longer independent since they
interact through instantaneous hard collisions. If the particles of
the background fluid, with mass $\mbf$, are much lighter than the
Brownian particles, with mass $m$, the drag force is usually assumed
to be linear in the velocity of the particles,
$\bm{F}_{\text{drag}}=-m\zeta_{0}\bm{v}$. In fact, this is the leading
behavior found when an expansion in powers of $\mbf/m$ is performed,
which leads to linear Brownian motion. Nevertheless, the drag force
becomes nonlinear when higher order terms in the expansion are brought
to bear. Specifically, the drag force can be written as
$\bm{F}_{\text{drag}}=-m\zeta(v)\bm{v}$ and there appears a
velocity-dependent drag coefficient $\zeta(v)$, with
$\zeta(v=0)=\zeta_{0}$. Therein, the first correction in $\zeta_{0}$
introduces a quadratic dependence on $v$.\cite{F07,F14,HKLMSLW17} In
some situations, nonlinearities in the drag coefficient have quite
strong physical implications.\cite{SAG95,F98,M13,HKLMSLW17,PWCNT18}

The main goal of this paper is to study the Mpemba effect in the
kinetic theory framework we have just described, i.e., the
Enskog--Fokker--Planck equation with nonlinear drag. To meet this end,
we work using the first Sonine approximation, in which the time evolution
of temperature is coupled to that of excess kurtosis. This
coupling, which is absent in the case $\zeta(v)=\zeta_0$, is responsible for the emergence of the Mpemba effect. The
value of  excess kurtosis is assumed to be small in the Sonine
approximation, which entails that the initial temperatures of the
samples must be close to each other and the Mpemba crossover takes place in the
early stage of relaxation. This allows us to
linearize the problem and derive
analytical expressions for the relevant physical quantities that
characterize the Mpemba effect, like the crossing time in the
temperature evolution, the maximum value of the initial temperature
difference, and the magnitude of the effect.

The plan of the paper is as follows. We put forward the model and the
kinetic description in Sec.~\ref{sec2}, where the equations for the
velocity moments are also derived. Section~\ref{sec3} is devoted to
the Sonine approximation: the infinite hierarchy for the velocity
moments is closed by expanding the VDF in Laguerre polynomials,
retaining only the first nontrivial cumulant, namely,  excess
kurtosis. The Mpemba effect is analyzed in Sec.~\ref{sec4}: we develop
a linearized model, investigate the crossover time, construct the
phase diagram in the space of parameters, quantify the magnitude of
the effect, and finally study the accuracy of the linearized
theory. Finally, Sec.~\ref{sec-concl} presents the main conclusions of
our work.

\section{Enskog--Fokker--Planck equation: Moment equations}
\label{sec2}

Let us consider a $d$-dimensional  system of elastic hard spheres of
mass $m$ and diameter $\sigma$ in a uniform and isotropic fluidized
state. The spheres are assumed to be suspended in a background fluid
in equilibrium so that their one-body VDF $f(\bm{v})$ satisfies the   Enskog--Fokker--Planck equation
\begin{equation}
\label{0}
\partial_t f(\bm{v})-\frac{\partial}{\partial \bm{v}}\cdot \left[\zeta(v)\bm{v}+\frac{\xi^2(v)}{2}\frac{\partial}{\partial \bm{v}} \right]f(\bm{v})
=J[\bm{v}|f,f].
\end{equation}
On the one hand, the force exerted by the background fluid on the
Brownian particles has two components: a nonlinear drag force
$\bm{F}_{\text{drag}}=-m\zeta(v)\bm{v}$ and a white-noise stochastic
force with nonlinear variance $m^2\xi^2(v)$. On the other hand,
collisions between Brownian particles are accounted for by the
Boltzmann--Enskog collision operator\cite{CC70,C88}
\bal
\label{6}
J[\bm{v}_1|f,f]=&\sigma^{d-1}g(\sigma)\int d\bm{v}_2\int d\widehat{\bm{\sigma}}\, \Theta(\bm{v}_{12}\cdot\widehat{\bm{\sigma}})\bm{v}_{12}\cdot\widehat{\bm{\sigma}}\nn
&\times[f(\bm{v}_1')f(\bm{v}_2')-f(\bm{v}_1)f(\bm{v}_2)].
\eal
Therein,
$g(\sigma)=\lim_{r\to\sigma^+}g(r)$ is the contact value of the pair correlation function $g(r)$,
$\Theta(\cdot)$ is the Heaviside step function, $\bm{v}_{12}\equiv \bm{v}_{1}-\bm{v}_{2}$ is the relative velocity, and
\begin{equation}
\bm{v}_1'=\bm{v}_1-(\bm{v}_{12}\cdot\widehat{\bm{\sigma}})\widehat{\bm{\sigma}},\quad \bm{v}_2'=\bm{v}_2+(\bm{v}_{12}\cdot\widehat{\bm{\sigma}})\widehat{\bm{\sigma}}
\end{equation}
are postcollisional velocities. Note that in the spatially uniform
states under consideration,  the Enskog collision operator \eqref{6} is simply Boltzmann's multiplied by the factor $g(\sigma)$.

The velocity-dependent coefficients $\zeta(v)$ and $\xi(v)$ are related by the condition that Eq.\ \eqref{0} admits as a stationary solution the  equilibrium VDF,
\begin{equation}
\label{18}
f_\st(\bm{v})=n\left(\frac{m}{2\pi k_B T_{\st}}\right)^{d/2} e^{-mv^2/2k_B T_{\st}},
\end{equation}
where $k_B$ is the Boltzmann constant and $T_{\st}$ is the equilibrium temperature of the background fluid, which acts as a thermostat.
This yields the fluctuation--dissipation relation
\begin{equation}
\xi^2(v)=\frac{2k_BT_{\st}}{m}\zeta(v).
\end{equation}

Equation~\eqref{0} can then be rewritten as
\begin{equation}
\label{1}
\partial_t f(\bm{v})-\frac{\partial}{\partial \bm{v}}\cdot
\zeta(v)\left(\bm{v} +\frac{k_BT_{\st}}{m}\frac{\partial}{\partial
    \bm{v}}\right)f(\bm{v}) =J[\bm{v}|f,f],
\end{equation}
which
describes the Brownian motion of an ensemble of particles of mass $m$
moving in the background fluid. These
Brownian particles are not independent, their interaction being
incorporated through the collision term.

In this work, as the simplest
nonlinear model, we consider the quadratic dependence of the drag
coefficient on the velocity derived in
Refs.~\onlinecite{F07,F14,HKLMSLW17}. Thus,
we restrict ourselves to
\begin{equation}
  \label{eq:zeta-v}
\zeta(v)=\zeta_0\left(1+\gamma \frac{m v^2}{k_BT_{\st}}\right),
\end{equation}
where $\gamma>0$ is a dimensionless parameter measuring the degree of
nonlinearity of the drag force.
When both Brownian particles  and background fluid particles are three-dimensional hard spheres,
  it has been shown that $\gamma=\mbf/10m$.\cite{F07,F14,HKLMSLW17}  See
  also Appendix \ref{app-B}. In this work, we restrict ourselves to $\mbf\leq 2m$, i.e., $\gamma\leq 0.2$.

By taking velocity moments in Eq.\ \eqref{1}, the evolution equation
for  temperature
\begin{equation}
\label{temperature}
T=\frac{m}{k_Bd}\langle v^2\rangle=\frac{m}{k_Bnd}\int d\bm{v}\, v^2 f(\bm{v}),
\end{equation}
where $n\!=\!\int d\bm{v} f(\bm{v})$ is the number density, is
obtained as
\begin{equation}
\label{25}
\frac{\dot{T}}{\zeta_0}=2\left(T_{\st}-T\right)\left[1+\gamma (d+2)\frac{T}{T_{\st}}\right]
-2\gamma(d+2)\frac{T^2}{T_{\st}}a_2,
\end{equation}
in which we have introduced the excess kurtosis
\begin{equation}
\label{4}
a_2=\frac{d}{d+2}\frac{\langle v^4\rangle}{\langle v^2\rangle^2}-1.
\end{equation}

In the particular case of a linear drag, $\gamma=0$, the solution to
Eq.\ \eqref{25} is simply
$T(t)=T_{\st}+\left[T(0)-T_{\st}\right]e^{-2\zeta_0 t}$.  However, in
the case of nonlinear drag, $\gamma>0$, the evolution of
temperature is coupled to that of excess kurtosis. Imagine that
$T(0)>T_{\st}$; the larger the value of $a_2(0)$, the larger the
initial cooling rate is and the sooner the temperature is expected to
reach the thermostat value $T_{\st}$. This property can give rise to an
Mpemba phenomenon, as reported in the case of granular
fluids.\cite{LVPS17} Similarly, the inverse Mpemba effect, in which
the cooler system heats sooner,\cite{LR17,LVPS17,BPRR20} may also be
expected for $T(0)<T_{\st}$.

Since the evolution equation \eqref{25} involves the excess kurtosis $a_2(t)$, we need to consider its evolution equation. This in turn involves sixth-degree moments, and so on, giving rise to an infinite hierarchy of moment equations. To derive this hierarchy,  let us introduce the dimensionless VDF $\phi(\bm{c})$ as
\begin{equation}
\label{5}
f(\bm{v})=\frac{n}{v_T^{d}(t)} \phi(\bm{c}),\quad \bm{c}\equiv\frac{\bm{v}}{v_T(t)},
\end{equation}
where $v_T(t)\equiv \sqrt{2k_BT(t)/m}$ is the thermal velocity.
Then, the kinetic equation \eqref{1} becomes
\begin{align}
\label{7}
\partial_t\phi(\bm{c})-\frac{\partial}{\partial \bm{c}}\cdot&\left[\frac{\dot{T}}{2T}\bm{c}+\zeta_{0}\left(1+\gamma\frac{2T}{T_{\st}} c^2\right)\right.
\nn
&\left.\;\times\!\!\left(\bm{c}+\frac{T_{\st}}{2T}\frac{\partial}{\partial \bm{c}}\right)\right]\phi(\bm{c})
=\nu_{\st} \sqrt{\frac{T}{T_{\st}}}I[\bm{c}|\phi,\phi],
\end{align}
in which we have defined $\nu_{\st}\equiv g(\sigma)
n\sigma^{d-1}\sqrt{2k_{B}T_{\st}/m}$, which is
basically the collision frequency at the steady state, and the
dimensionless collision operator
\bal
\label{8}
I[\bm{c}_1|\phi,\phi]=&\int  d\bm{c}_2\int d\widehat{\bm{\sigma}}\, \Theta(\bm{c}_{12}\cdot\widehat{\bm{\sigma}})\,\bm{c}_{12}\cdot\widehat{\bm{\sigma}}\nn
&\times
[\phi(\bm{c}_1')\phi(\bm{c}_2')-\phi(\bm{c}_1)\phi(\bm{c}_2)].
\eal
Also, we have employed the property
\begin{equation}
\label{9}
\frac{v_T^d}{n}\partial_t f(\bm{v})=\partial_t \phi(\bm{c})-\frac{\dot{T}}{2T}\frac{\partial}{\partial \bm{c}}\cdot \left[\bm{c}\phi(\bm{c})\right].
\end{equation}

Multiplying both sides of Eq.\ \eqref{7} by $c^\ell$, integrating over
$\bm{c}$, and making use of Eq.\ \eqref{25}, we obtain the
hierarchy of equations for the moments
$M_\ell\equiv \langle c^\ell\rangle$,
\begin{align}
\label{28}
\dot{M}_\ell =&\ell\zeta_{0}\left\{
\left[\gamma(\ell-2)+\gamma(d+2)\frac{T}{T_{\st}}(1+a_2)-\frac{T_{\st}}{T}\right]M_\ell\right.
\nn &-\left.\gamma\frac{2 T}{T_{\st}} M_{\ell +2}+ \frac{\ell
  +d-2}{2}\frac{T_{\st}}{T}M_{\ell
  -2}\right\}-\nu_{\st}\sqrt{\frac{T}{T_{\st}}}\mu_\ell .
\end{align}
Here, we have introduced the collisional moments $\mu_{\ell}$ as
\begin{equation}
\label{12}
\mu_\ell \equiv -\int d\bm{c}\, c^\ell  I[\bm{c}|\phi,\phi].
\end{equation}
Since, by definition, $M_0=1$, $M_2=d/2$, and $M_4=d(d+2)(1+a_2)/4$ [see Eqs.\ \eqref{temperature} and \eqref{4}], it is easy to check that, as it should be, $\dot{M}_0=\dot{M}_2=0$ (note that $\mu_0=\mu_2=0$).
Next, setting $\ell=4$ in Eq.\ \eqref{28}, we get
\begin{align}
\label{14}
\dot{a}_2=&\zeta_{0}\gamma\frac{4T}{T_{\st}}\Bigg[\frac{2T_{\st}}{T}(1+a_2)+(d+2)(1+a_2)^2-(d+4)\nn
&\times(1+3a_2-a_3)\Bigg]-\zeta_{0}\frac{4T_{\st}}{T}a_2-\frac{4\nu_{\st}}{d(d+2)}\sqrt{\frac{T}{T_{\st}}}\mu_4,
\end{align}
where we have introduced  the sixth-degree cumulant $a_3$ by
\begin{equation}
\label{a3}
M_6=\frac{d(d+2)(d+4)}{8}(1+3a_2-a_3).
\end{equation}

Some comments are in order. First, note that two time scales compete
in Eqs.\ \eqref{28} and \eqref{14}. The inverse of the drag
coefficient for low velocities, $\zeta_0^{-1}$, dictates the time
scale over which particles feel the action of the background
fluid. Meanwhile, the characteristic time for particle--particle
collisions is the inverse of the stationary collision frequency,
$\nu_{\st}^{-1}$. Second, Equations \eqref{25}, \eqref{28}, and
\eqref{14} are formally exact within the Enskog--Fokker--Planck
description, but they do not make a closed finite set. Not only does
$\dot{M_\ell}$ explicitly involve a higher-degree moment $M_{\ell+2}$
but also the collisional moment $\mu_\ell$ is a nonlinear functional
of the full VDF $\phi(\bm{c})$. An approximate closure is
needed to deal with a finite set of equations.

\section{Sonine approximation}\label{sec3}

For isotropic states, the reduced VDF $\phi(\bm{c})$ can be expanded in a complete set of orthogonal polynomials as
\begin{equation}
\label{Sonine}
\phi(\bm{c})=\frac{e^{-c^2}}{\pi^{d/2}}\left[1+\sum_{\ell=2}^\infty
  a_\ell L_\ell^{(\frac{d-2}{2})}(c^2)\right], \end{equation}
where
$L_\ell^{(\alpha)}(x)$ are generalized Laguerre (or Sonine)
polynomials.\cite{AS72} Of course, the coefficients with $\ell=2$ and $\ell=3$ in Eq.\ \eqref{Sonine} are the same as the cumulants $a_2$ and $a_3$, respectively, introduced before [see Eqs.\ \eqref{4} and \eqref{a3}].

In the first Sonine approximation, all terms
beyond $\ell=2$ in Eq.~\eqref{Sonine} are dropped, i.e.,
\begin{equation}
\label{1stSonine}
\phi(\bm{c})\approx\frac{e^{-c^2}}{\pi^{d/2}}\left\{1+ a_2 \left[\frac{c^4}{2}-\frac{d+2}{2}c^2+\frac{d(d+2)}{8}\right]\right\}.
\end{equation}
Inserting Eq.\ \eqref{1stSonine} into Eq.\ \eqref{12} with $\ell=2$
and neglecting terms quadratic in $a_2$, one obtains $\Gamma(d/2)\mu_4\approx \sqrt{2}(d-1)\pi^{\frac{d-1}{2}}a_2$.\cite{vNE98,MS00,SM09}
Therefore, Eq.\ \eqref{14} becomes
\begin{align}
\label{16}
\dot{a}_2=&8\zeta_{0}\gamma\left(1-\frac{T}{T_{\st}}\right)-\left\{
\zeta_{0}\left[ \frac{4T_{\st}}{T}-8\gamma+4\gamma(d+8)
\frac{T}{T_{\st}}\right]\right.\nn
&\left.+\tau_{\st}^{-1}\frac{8(d-1)}{d(d+2)}
\sqrt{\frac{T}{T_{\st}}}\right\}a_2,            ,
\end{align}
where we have introduced the mean free time at the steady state\cite{RL77,PL01}
\begin{equation}
  \tau_\st=\sqrt{2}\,\Gamma(d/2)
  \pi^{\frac{1-d}{2}}\nu_{\st}^{-1},
\end{equation}
and, for consistency, the terms $a_2^2$ and $a_3$ have been neglected.

Equations \eqref{25} and \eqref{16} make a closed set to investigate
the existence of the Mpemba effect. First, we
define dimensionless temperature and time by
\begin{equation}
\theta\equiv\frac{T}{T_{\st}}, \qquad
t^*\equiv \frac{t}{\tau_\st}.
\end{equation}
The latter approximately measures the accumulated number of collisions
per particle up to time $t$. With these variables, Eqs.~\eqref{25} and \eqref{16} can be rewritten as
\begin{subequations}
\label{25b-30}
\bal
\label{25b}
\frac{\dot{\theta}}{\zeta_0^*}=&2\left(1-\theta\right)\left[1+\gamma (d+2)\theta\right]
-2\gamma(d+2){\theta^2}a_2,\\
\label{30}
\frac{\dot{a}_2}{\zeta_0^*}=&8\gamma\left(1-\theta\right)-\left[\frac{4}{\theta}-8\gamma+4\gamma(d+8)\theta
  +\frac{8(d-1)}{d(d+2)}\frac{\sqrt{\theta}}{\zeta_0^*}\right]
  \nn&\times a_2,
\eal
\end{subequations}
where we have introduced a dimensionless low-velocity drag coefficient
as
\begin{equation}
  \label{eq:z0*}
  \zeta_0^*\equiv \zeta_0\tau_{\st}.
\end{equation}
Now, the dot over $\theta$ and $a_2$ denotes a derivative with
respect to $t^*$.

Equations~\eqref{25b-30} are linear in the excess kurtosis $a_2$ but
nonlinear in the temperature ratio $\theta$. They constitute our
starting point for the analysis of the Mpemba effect, to be carried
out in Sec.\ \ref{sec4}. In the dimensionless variables we are using,
there are only two relevant parameters: (i) $\gamma$, which measures
the strength of the nonlinearity in the drag, and (ii)
$\zeta_{0}^{*}$, which compares the characteristic times for
collisions, $\tau_{\st}$, and for the viscous drag,
$\zeta_{0}^{-1}$. Note that the regime $\zeta_{0}^{*}\ll 1$
($\zeta_{0}^{*}\gg 1$) means that the viscous drag acts over a much
longer (shorter) time scale than collisions do.

\section{Mpemba effect}
\label{sec4}

\subsection{Linearized model}

Let us imagine two initial states A and B with
$\{\theta(0),a_{2}(0)\}=\{\theta_{A}^0,a_{2A}^0\}$ and $\{\theta_{B}^0,a_{2B}^0\}$,
respectively.  The corresponding solutions to Eqs.\ \eqref{25b-30} are denoted by $\{\theta_A(t^*),a_{2A}(t^*)\}$ and $\{\theta_B(t^*),a_{2B}(t^*)\}$.
Without loss of generality, we assume that
$\theta_{A}^0>\theta_{B}^0$.\footnote{Throughout, the superscript $0$ denotes
  initial value for all the variables.}

Below, we show that both the Mpemba effect and its inverse version are
expected to emerge when the initially hotter sample has the larger
value of the excess kurtosis. First, we analyze the case in which both
initial temperatures are higher than the stationary one,
$\theta_{A}^0>\theta_{B}^0>1$, and the system cools down to reach the
steady state. The Mpemba effect is present when $\theta_{A}(t^*)$
relaxes more rapidly than $\theta_{B}(t^*)$, which calls for the
existence of a crossover time $t_c^*$ such that
$\theta_A(t_{c}^*)=\theta_B(t_{c}^*)$. Since the cooling rate increases with
$a_2$, the condition $a_{2A}^0>a_{2B}^0$ seems to be necessary for the
Mpemba effect to emerge. Second, we look into the case in which both
temperatures are lower than the stationary value,
$1>\theta_{A}^0>\theta_{B}^0$, and the system heats up. The inverse
Mpemba effect appears if $\theta_{A}(t^*)$ relaxes more slowly than
$\theta_{B}(t^*)$, which again needs that $a_{2A}^0>a_{2B}^0$.

\begin{figure*}
\includegraphics[width=\ancho]{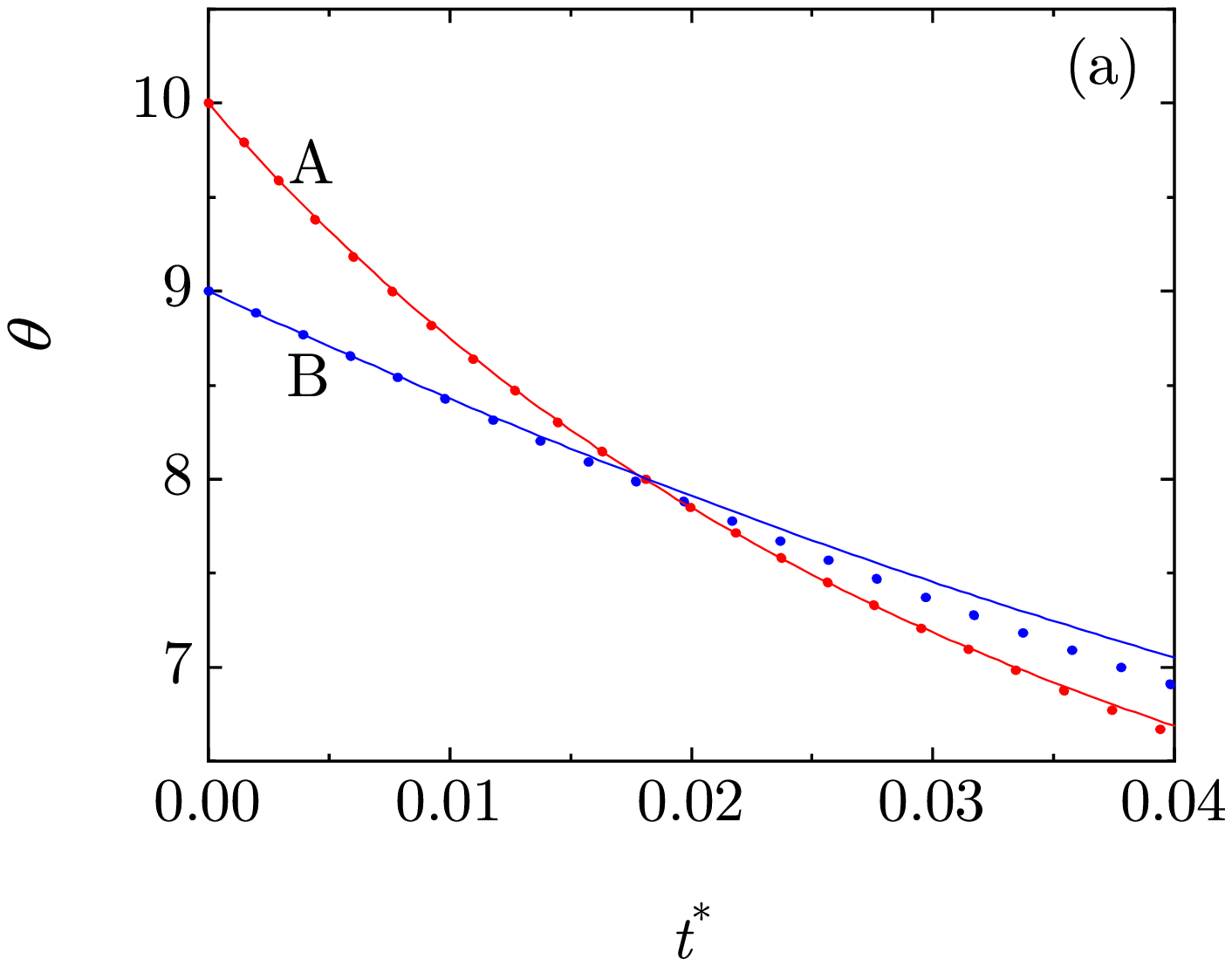}
\includegraphics[width=\ancho]{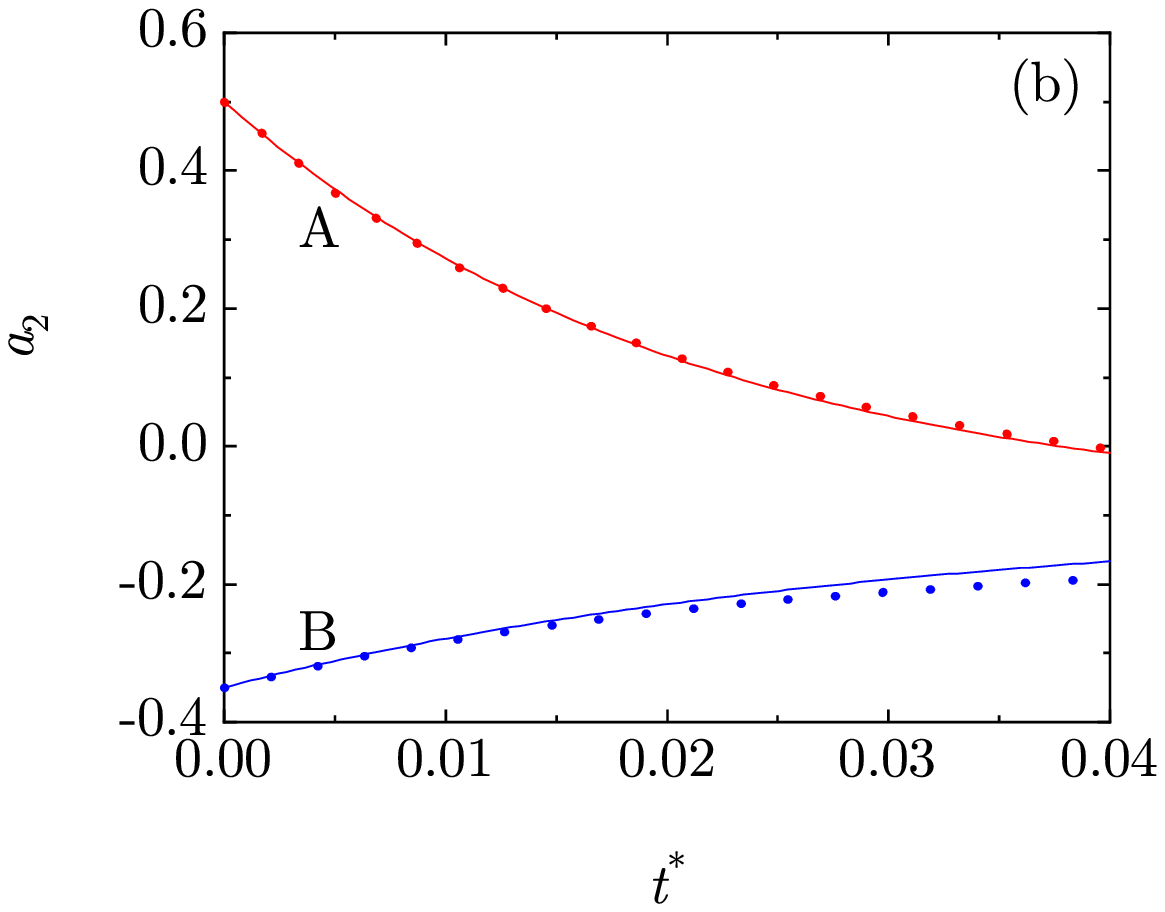}\\
\vspace{2mm}
\includegraphics[width=\ancho]{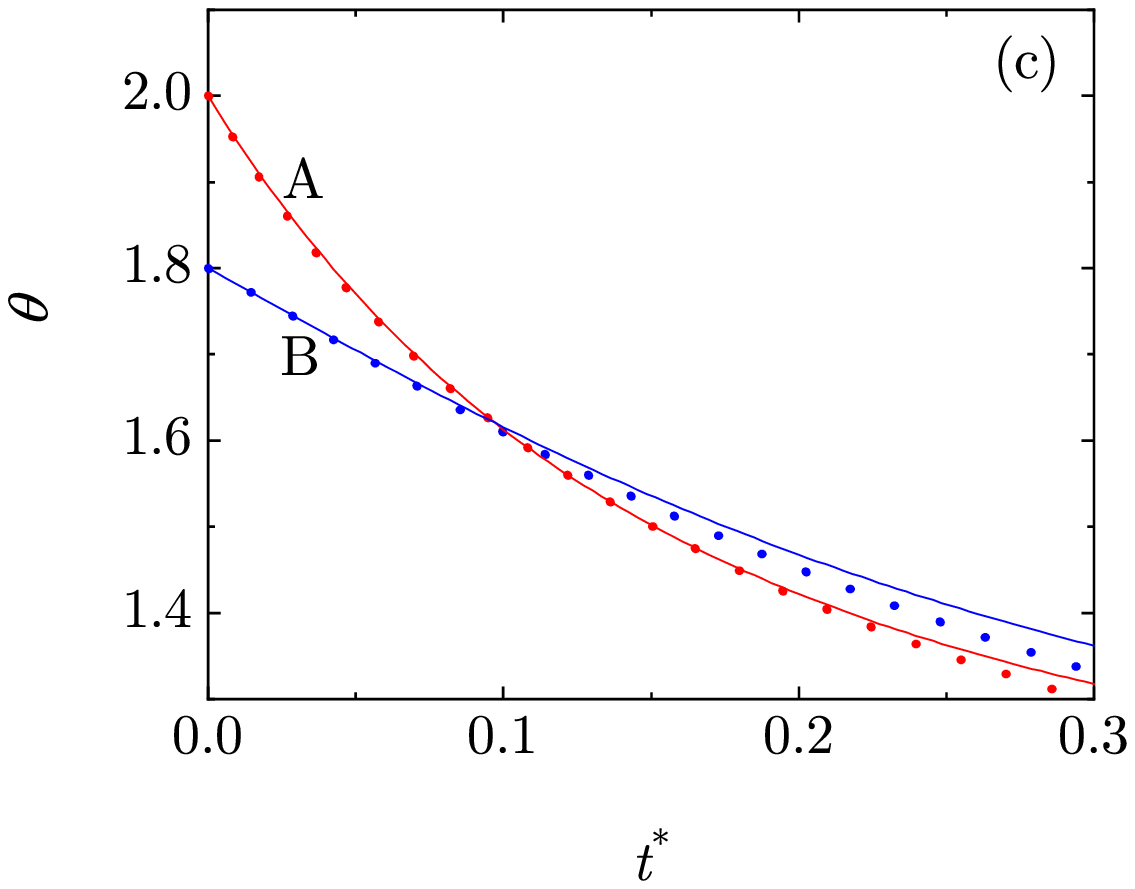}
\includegraphics[width=\ancho]{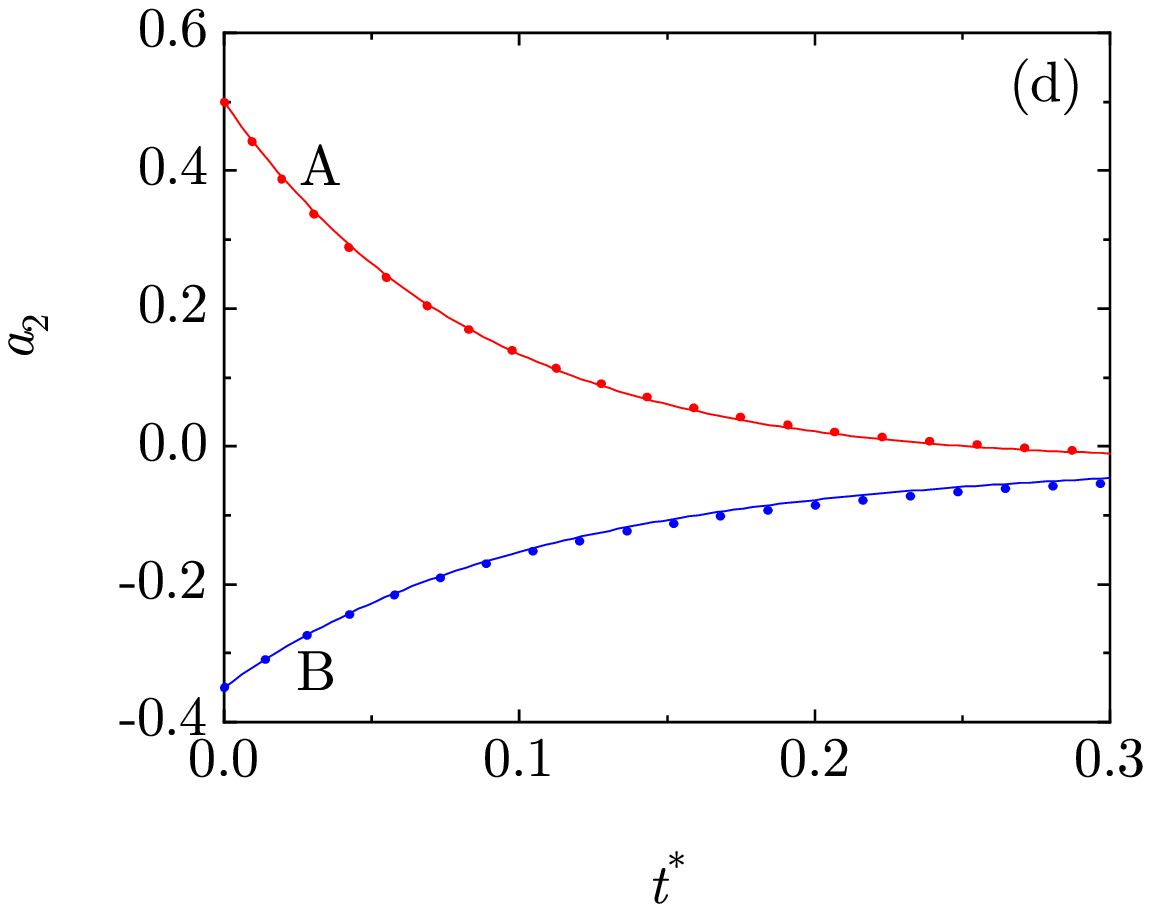}\\
\vspace{2mm}
\includegraphics[width=\ancho]{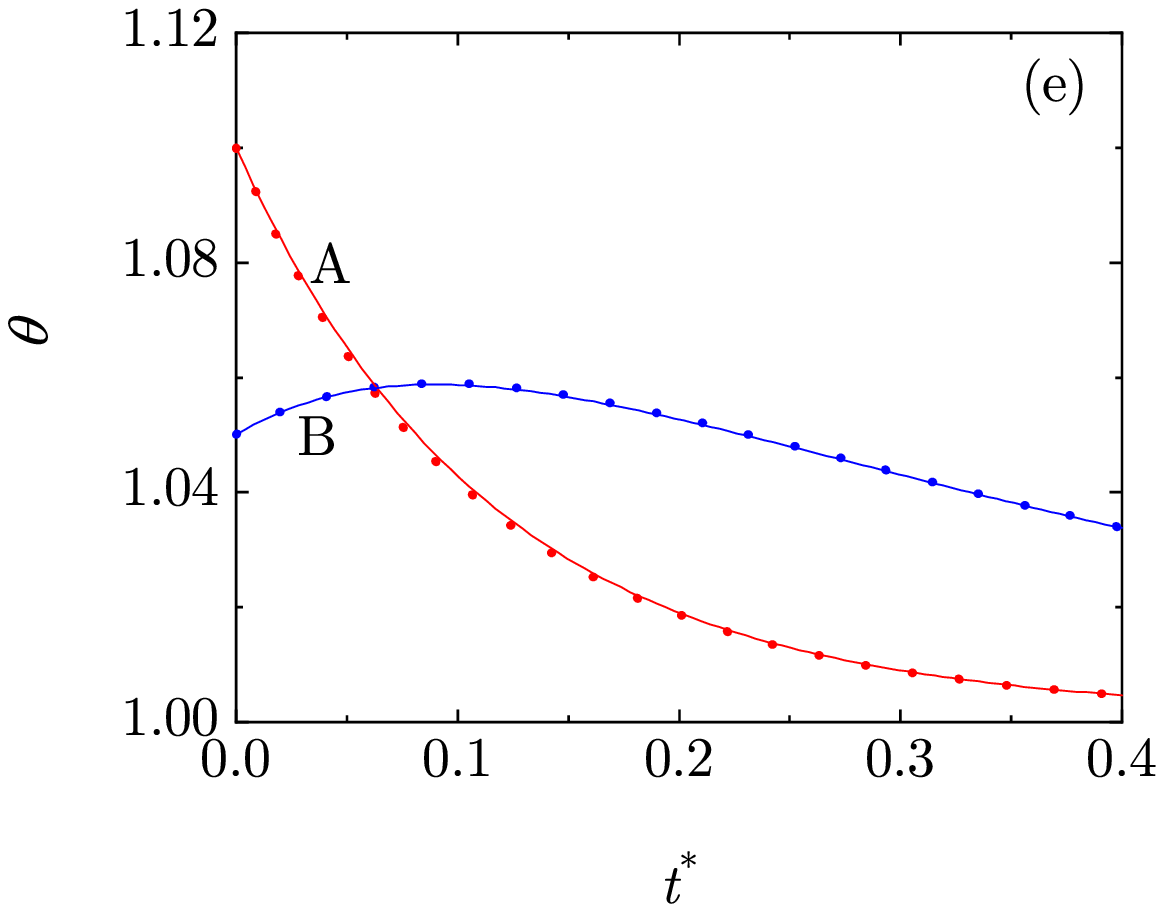}
\includegraphics[width=\ancho]{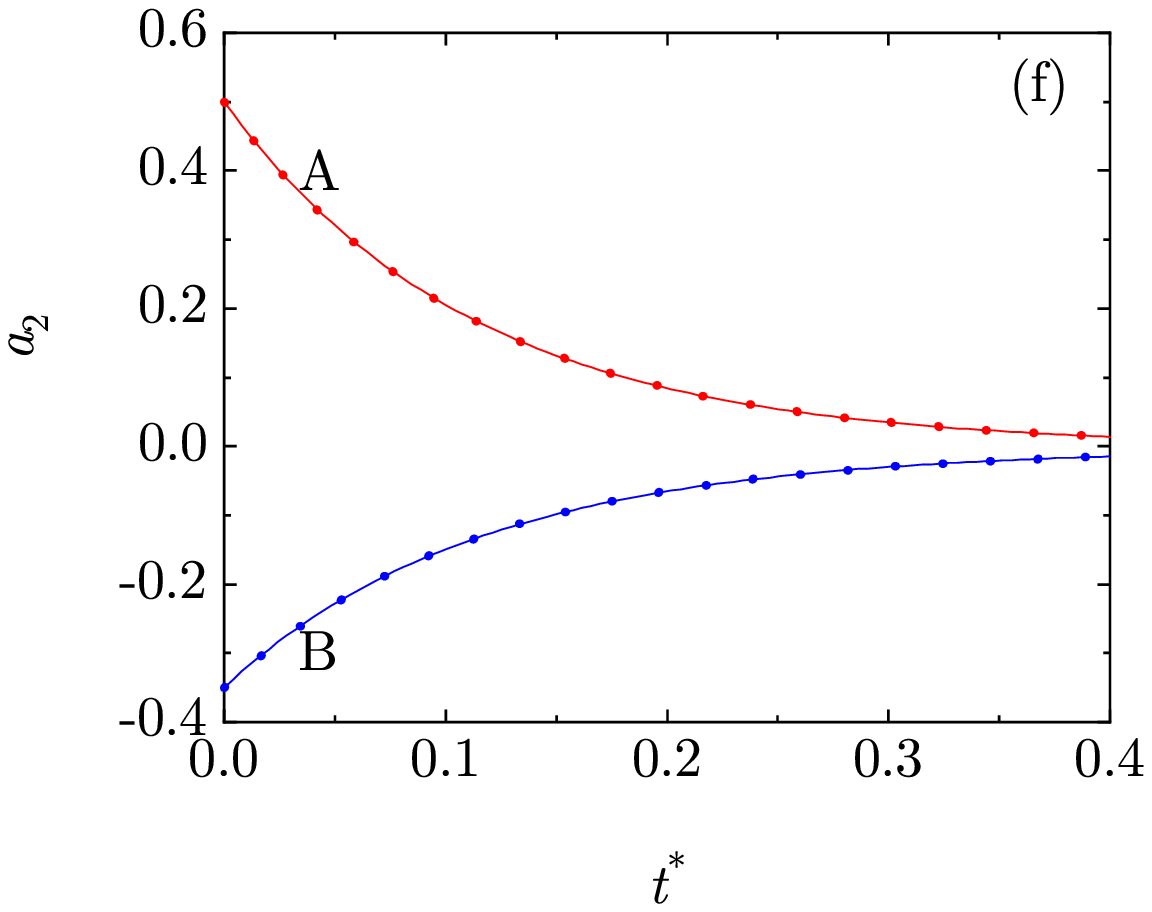}\\
\vspace{2mm}
\includegraphics[width=\ancho]{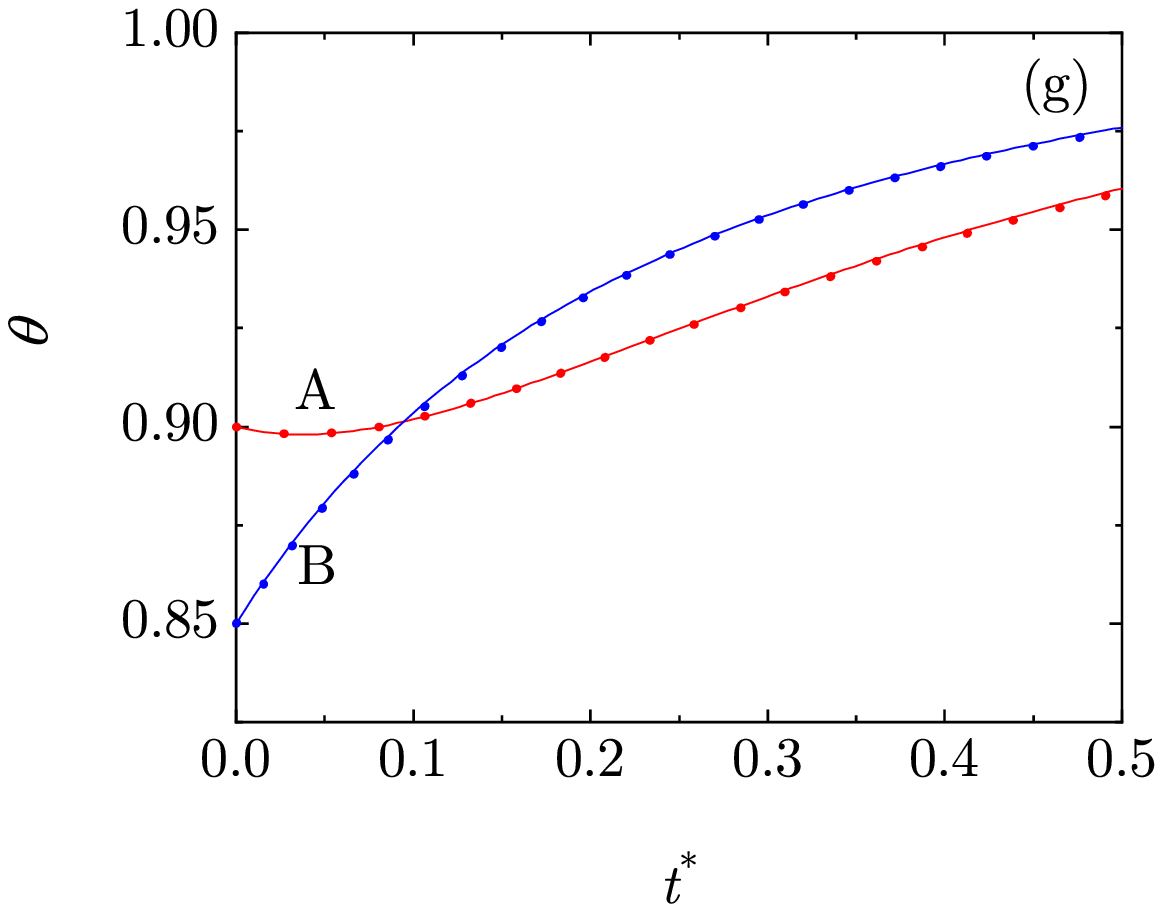}
\includegraphics[width=\ancho]{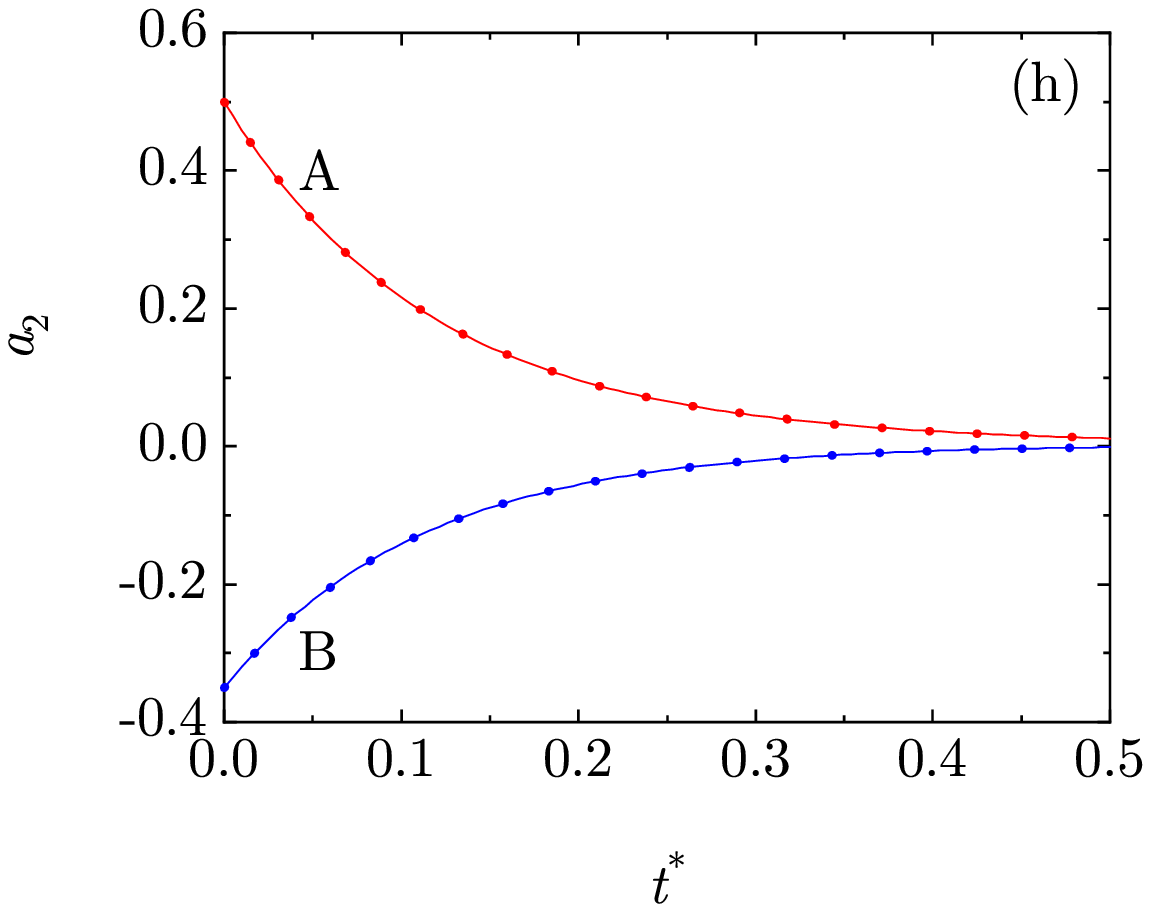}
\caption{\label{fig1} First stage in the evolution of $\theta_A(t^*)$
  and $\theta_B(t^*)$ [panels (a), (c), (e), and (g)] and $a_{2A}(t^*)$
  and $a_{2B}(t^*)$ [panels (b), (d), (f), and (h)] for $d=3$, $\zeta_0^*=1$, and $\gamma=0.1$. The
  initial states are $a_{2A}^0=0.5$, $a_{2B}^0=-0.35$, and  [(a) and (b)] $\theta_A^0=10$, $\theta_B^0=9$,
  [(c) and (d)] $\theta_A^0=2$, $\theta_B^0=1.8$, [(e) and (f)] $\theta_A^0=1.1$, $\theta_B^0=1.05$, and [(g) and (h)] $\theta_A^0=0.9$, $\theta_B^0=0.85$. Circles correspond to
  the numerical solutions of Eqs.~\eqref{25b-30}, whereas solid lines correspond to
  the linearized model, given by Eqs.~\eqref{relax-explicit}. The
  Mpemba effect is neatly observed in panels (a), (c), and (e). The
  linearized theory with $\theta_r=\theta_{B}^0$ gives a correct
  account thereof---although it deviates from the numerical solution
  of Eqs.~\eqref{25b-30} as time increases in panels (a) and (c), for
  which their initial temperatures are not close to the steady
  value. The inverse Mpemba effect is depicted in panel (g) and the
  linear theory also describes it correctly, but now we have chosen
  $\theta_r=\theta_{A}^0$.}
\end{figure*}

In general, the nonlinear dependence on $\theta$ of the set of
equations \eqref{25b-30} impede a fully analytical treatment. However,
the excess kurtosis is supposed to be small in the Sonine
approximation, which has allowed us to neglect nonlinear terms in
$a_{2}$. Since $a_{2}$ is the quantity controlling the appearance of
the Mpemba effect, its smallness implies that both initial
temperatures, $\theta_{A}^0$ and $\theta_{B}^0$, cannot be very far from each other
for the Mpemba effect to emerge. In addition, the crossing of the
curves $\theta_{A}(t^{*})$ and $\theta_{B}(t^{*})$ must take place in
  the early stage of evolution.\footnote{A similar observation was
    found in Ref.~\onlinecite{LVPS17} for a granular fluid of smooth
    hard particles}

Following the discussion above, we write
$\theta(t^*)=\theta_r+\Psi(t^*)$, where
$\theta_r\approx \theta_{A}^0\approx\theta_{B}^0$ is a certain
\emph{reference} temperature, and linearize Eqs.\ \eqref{25b-30} with
respect to $\Psi(t^*)$ and $a_2(t^*)$. The detailed solution of this
linearization procedure is carried out in Appendix~\ref{app-A}; here, we
present the results relevant for the analysis of the Mpemba
effect. The time evolution is controlled by the matrix $\mathsf{\Lambda}$
with elements
\begin{subequations}
\label{4.3}
\bal
\label{4.3a}
{\Lambda}_{11}=&2\zeta_0^*\left[1+\gamma(d+2)\left(2\theta_r-1\right)\right],\\
\label{4.3b}
\Lambda_{12}=& 2\zeta_0^*\gamma(d+2)\theta_r^2,\\
\label{4.3c}
{\Lambda}_{21}=&8\zeta_0^*\gamma,\\
\label{4.3d}
{\Lambda}_{22}=&\zeta_0^*\left[\frac{4}{\theta_r}-8\gamma+4\gamma(d+8)\theta_r\right]
+\frac{8(d-1)}{d(d+2)}{\sqrt{\theta_r}},
\eal
\end{subequations}
and eigenvalues
\begin{equation}
\label{4.5}
\lambda_\pm=\frac{\Lambda_{11}+\Lambda_{22}\pm\delta_\lambda}{2},\quad
\delta_\lambda\equiv\sqrt{(\Lambda_{11}-\Lambda_{22})^2+4\Lambda_{12}\Lambda_{21}}.
\end{equation}

Let us consider the differences $\Delta \theta(t^*)\equiv
\theta_A(t^*)-\theta_B(t^*)$ and $\Delta a_2(t^*)\equiv
a_{2A}(t^*)-a_{2B}(t^*)$ between the time evolutions corresponding to
the two different initial states A and B. Within the linearized theory, these differences are given by (see Appendix~\ref{app-A})
\begin{subequations}
\label{relax-explicit2}
\bal
\label{temp-relax-explicit2}
\Delta \theta(t^*)=&\left(\frac{\lambda_{+}-\Lambda_{11}}{\delta_\lambda}\Delta \theta^0-
\frac{\Lambda_{12}}{\delta_\lambda} \Delta a_2^0\right)e^{-\lambda_{-}t^*}\nn
  &
-\left(\frac{\lambda_{-}-\Lambda_{11}}{\delta_\lambda}\Delta \theta^0-
\frac{\Lambda_{12}}{\delta_\lambda} \Delta a_2^0\right)e^{-\lambda_{+}t^*},\\
\label{a2-relax-explicit2}
\Delta a_2(t^*)=&\left(\frac{\lambda_{+}-\Lambda_{22}}{\delta_\lambda}\Delta a_2^0-
\frac{\Lambda_{21}}{\delta_\lambda} \Delta \theta^0\right)e^{-\lambda_{-}t^*}\nn
  &
-\left(\frac{\lambda_{-}-\Lambda_{22}}{\delta_\lambda}\Delta a_2^0-
\frac{\Lambda_{21}}{\delta_\lambda} \Delta \theta^0\right)e^{-\lambda_{+}t^*}.
\eal
\end{subequations}
Note that both $\Delta\theta$ and $\Delta a_2$ vanish in the long-time
limit.

\subsection{Mpemba crossover}

The accuracy of the linearized theory developed above for describing
the Mpemba effect---and also the inverse Mpemba effect---is illustrated in Fig.\ \ref{fig1}. The linear
theory remains valid even when the system is initially far from the
steady state; the
analytical expressions of the linearized theory, given by
Eqs.~\eqref{relax-explicit}, predict the crossover of the curves
correctly but start to deviate from the ``exact'' numerical
integration as time grows. For all these
plots, an optimal choice for $\theta_r$ is the initial temperature of
the sample that is closer to the steady state.
Figure \ref{fig1} also shows that the excess kurtosis relaxes to equilibrium more rapidly than temperature.

\begin{figure*}
\includegraphics[width=\ancho]{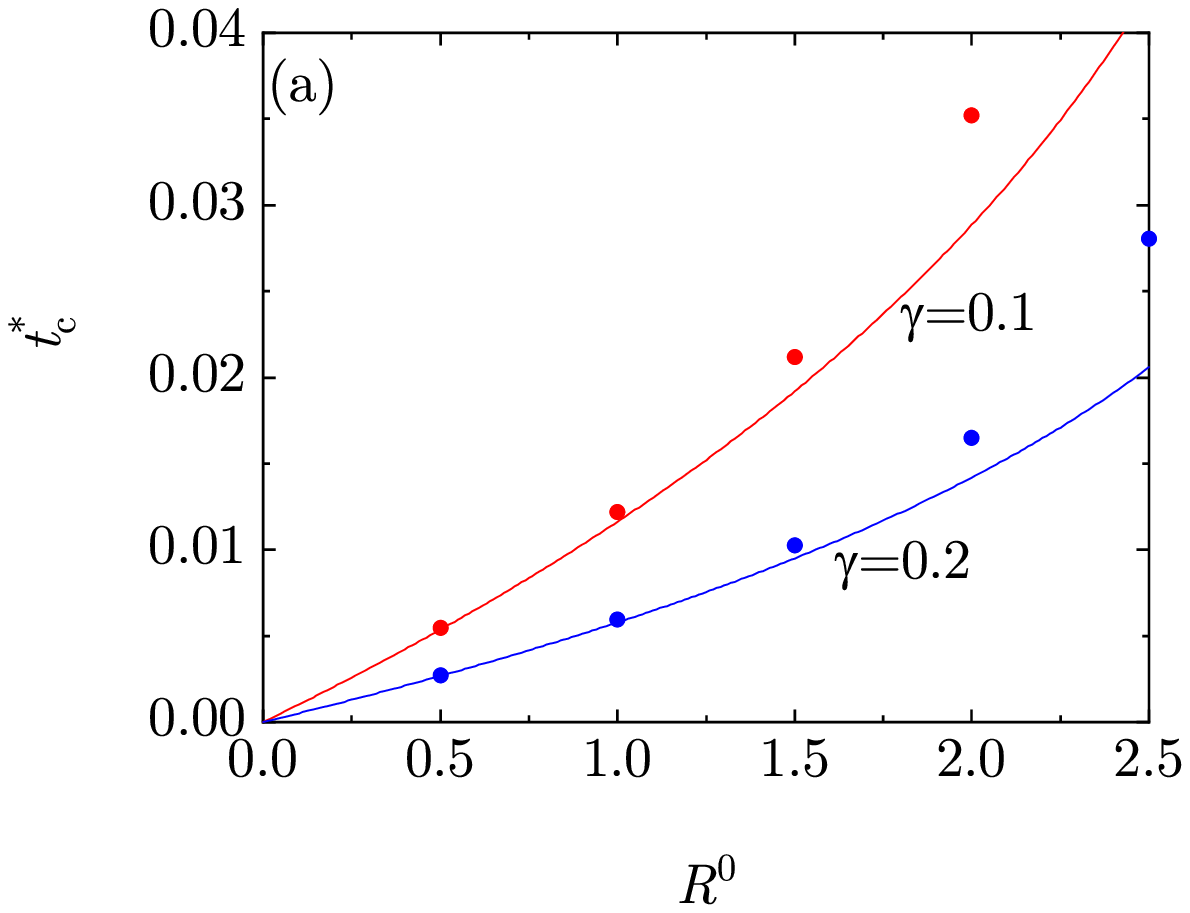}
\includegraphics[width=\ancho]{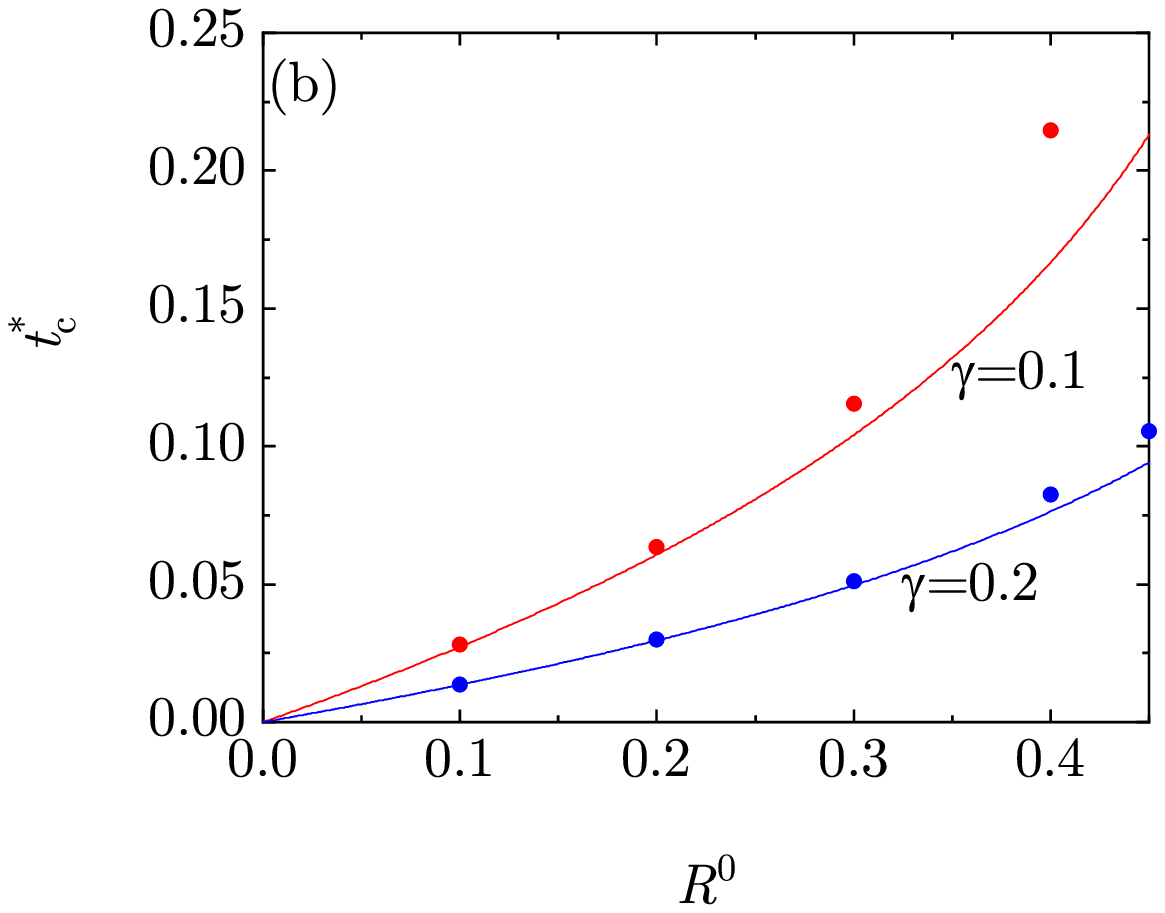}\\
\vspace{2mm}
\includegraphics[width=\ancho]{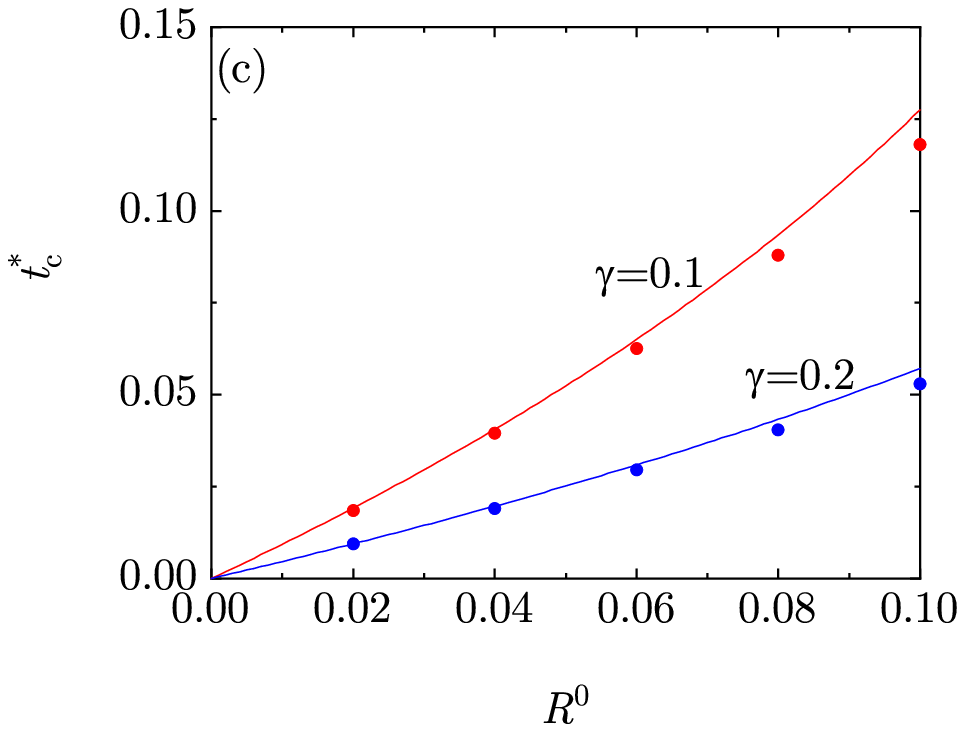}
\includegraphics[width=\ancho]{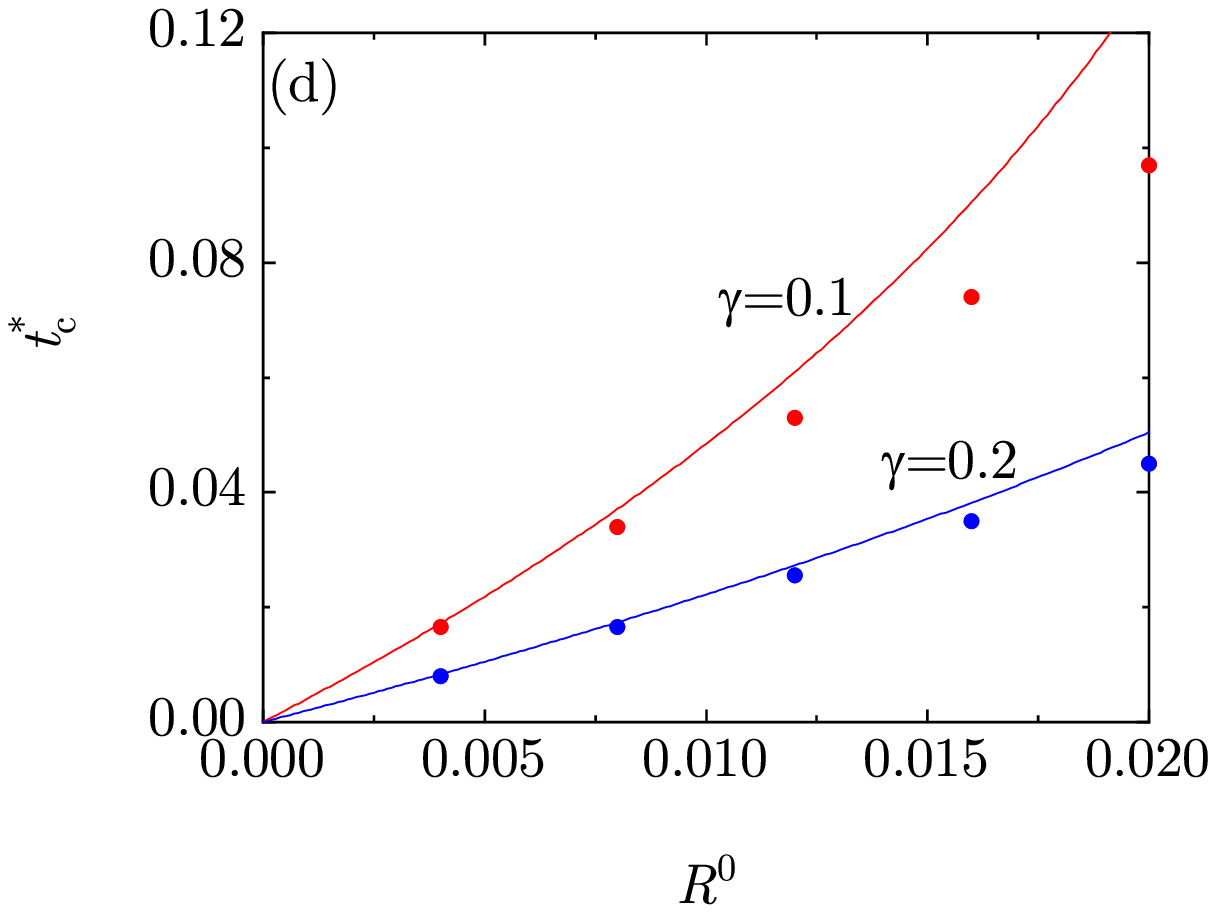}
\caption{\label{fig:t_c} Crossover time $t_c^*$ as a function of
  $R^{0}\equiv \Delta\theta^0/\Delta a_2^0$ [see Eq.\ \eqref{crossover-time}]. All the panels correspond to $d=3$ and $\zeta_0^*=1$, but
  to different values of the reference temperature: $\theta_r=10$,
  $2$, $1.05$, and $0.5$, from (a) to (d). In each panel, two values of
  the nonlinearity parameter are considered: $\gamma=0.1$ (upper
  curves) and $0.2$ (lower curves).
The crossover time decreases as the nonlinearity coefficient  $\gamma$ increases, as expected. The symbols are the values obtained from the numerical solution of Eqs.\ \eqref{25b-30} with $a_{2A}^0=0.5$ and $a_{2B}^0=-0.35$.}
\end{figure*}

Let us  now restrict ourselves to a situation in which the Mpemba
effect is present. Thus, there
exists a crossover time such that $\Delta \theta(t_c^*)=0$. According
to Eq.~\eqref{temp-relax-explicit2}, it is given by
\begin{equation}
\label{crossover-time}
t_{c}^*= \frac{1}{\delta_\lambda}\ln
\frac{\Lambda_{12}-(\lambda_{-}-\Lambda_{11})R^{0}}{\Lambda_{12}-(\lambda_{+}-\Lambda_{11})R^{0}}, \quad R^{0}\equiv\frac{\Delta\theta^{0}}{\Delta
  a_{2}^{0}}.
\end{equation}
The crossover time $t_c^*$ depends on the initial preparation
\emph{only} through the reference temperature
$\theta_r\approx \theta_{A}^0\approx\theta_{B}^0$ and the ratio
$R^{0}$ in this simplified description, for given values of
$\zeta_0^*$ and $\gamma$. Note that we have chosen
$\Delta\theta^{0}>0$ and, for the Mpemba effect to exist, we need that
$\Delta a_{2}^{0}>0$, i.e., we have that $R^{0}>0$.

Figure \ref{fig:t_c} displays $t_c^*$ as a function of $R^{0}$ for
some illustrative cases. Different panels correspond to different
values of the reference temperature. In all of them, the crossover
time $t_{c}^{*}$ vanishes in the limit as $R^{0}\to 0$ and grows with $R^{0}$.
Figure \ref{fig:t_c} also includes the values of the crossover time
obtained from the numerical solution of Eqs.\ \eqref{25b-30} for
$a_{2A}^0=0.5$ and $a_{2B}^0=-0.35$ with $\theta_B^0=\theta_r$ in
panels (a)--(c) and $\theta_A^0=\theta_r$ in panel (d). It is observed
that the agreement with Eq.\ \eqref{crossover-time} improves as
$\gamma$ increases and $R^0$ decreases. Also, Eq.\ \eqref{crossover-time} underestimates the crossover time for the direct Mpemba effect with initial temperatures far from that of the thermostat, while it tends to overestimate $t_c^*$ for the inverse Mpemba effect or when the initial temperatures are close to the thermostat one.

Equation \eqref{crossover-time} shows that $t_c^*$ diverges in
the limit as $R^{0}\to R^{0}_{\thr}$, where $R^{0}_{\thr}$ is a
\textit{threshold} value for the ratio, given by
\begin{equation}
\label{Delta-max}
R^{0}_{\thr}=
\frac{\Lambda_{12}}{\lambda_{+}-\Lambda_{11}}.
\end{equation}
Thus,
the Mpemba effect disappears if
$R^{0}\geq R^{0}_{\thr}$: in this region, $t_{c}^{*}$, as defined by
Eq.~\eqref{crossover-time}, ceases to be a real number. In fact, if we
define
\begin{equation}
  \label{eq:mu}
  \beta\equiv
  \frac{\lambda_{-}-\Lambda_{11}}{\lambda_{+}-\Lambda_{11}}=1-
  \frac{\delta_\lambda}{\lambda_{+}-\Lambda_{11}},
\end{equation}
we can rewrite $t_{c}^{*}$ as
\begin{equation}
  t_{c}^{*}=\frac{1}{\delta_\lambda} \ln \frac{1-\beta
    R^{0}/R^{0}_{\thr}}{1-R^{0}/R^{0}_{\thr}}  .
\end{equation}

The emergence of the Mpemba effect is basically controlled by the
strength of the drag nonlinearity $\gamma$.  As expected on a physical
basis, the Mpemba crossover takes place earlier as $\gamma$
increases. Throughout Fig.~\ref{fig:t_c}, the curves for
$\gamma=0.1$ lie above those for $\gamma=0.2$. Thus, the smaller
$\gamma$ is, the smaller the threshold
value $R^{0}_{\thr}$ we find. Recall that the drag
becomes linear in the limit as $\gamma\to 0^{+}$, for which the
temperature obeys a closed first-order differential
equation---independently of the value of the excess kurtosis, and the
Mpemba effect is no longer present.
Note also that terms beyond the quadratic one in the expansion of $\zeta(v)$ in powers of $v$ might be necessary  as $\gamma$ (or, equivalently, $\mbf/m$) increases.

\subsection{Phase diagram}

\begin{figure}
\includegraphics[width=\ancho]{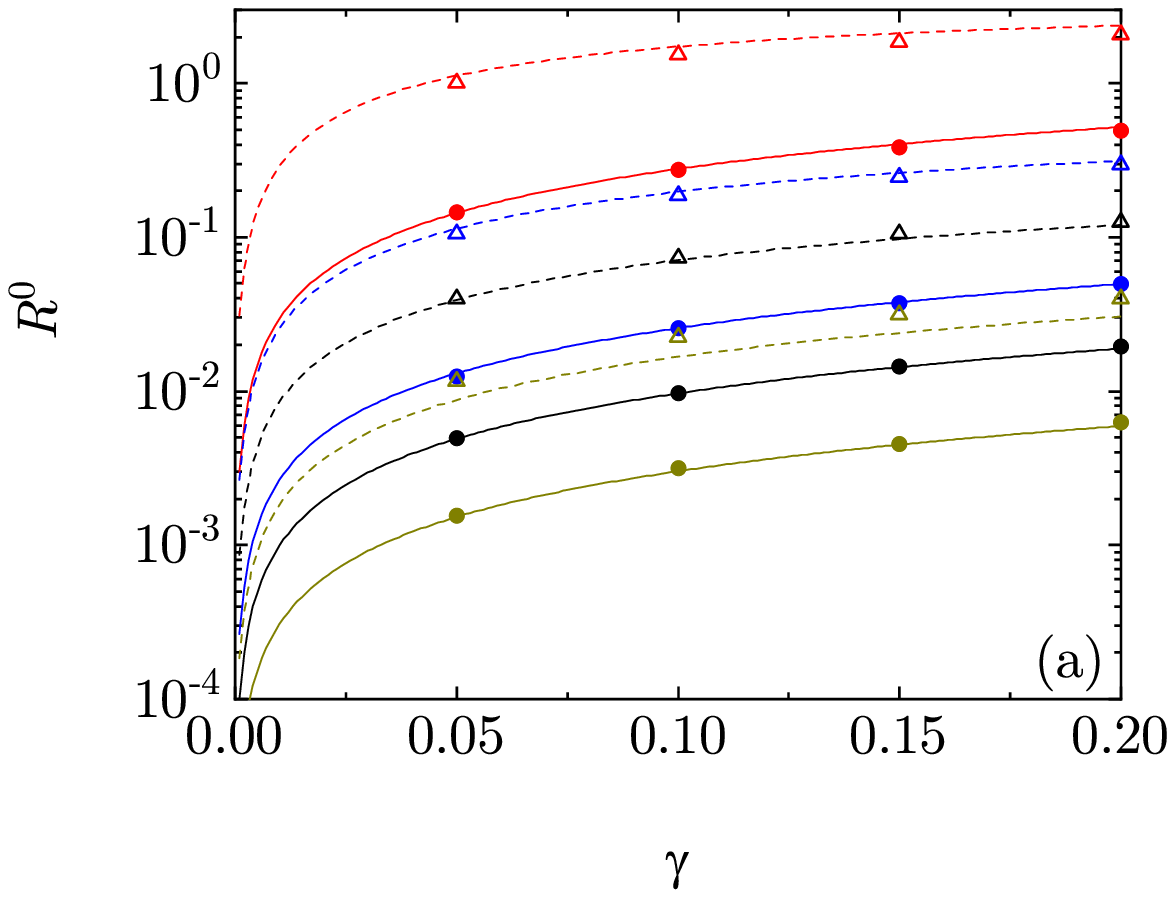}\\
\vspace{2mm}
\includegraphics[width=\ancho]{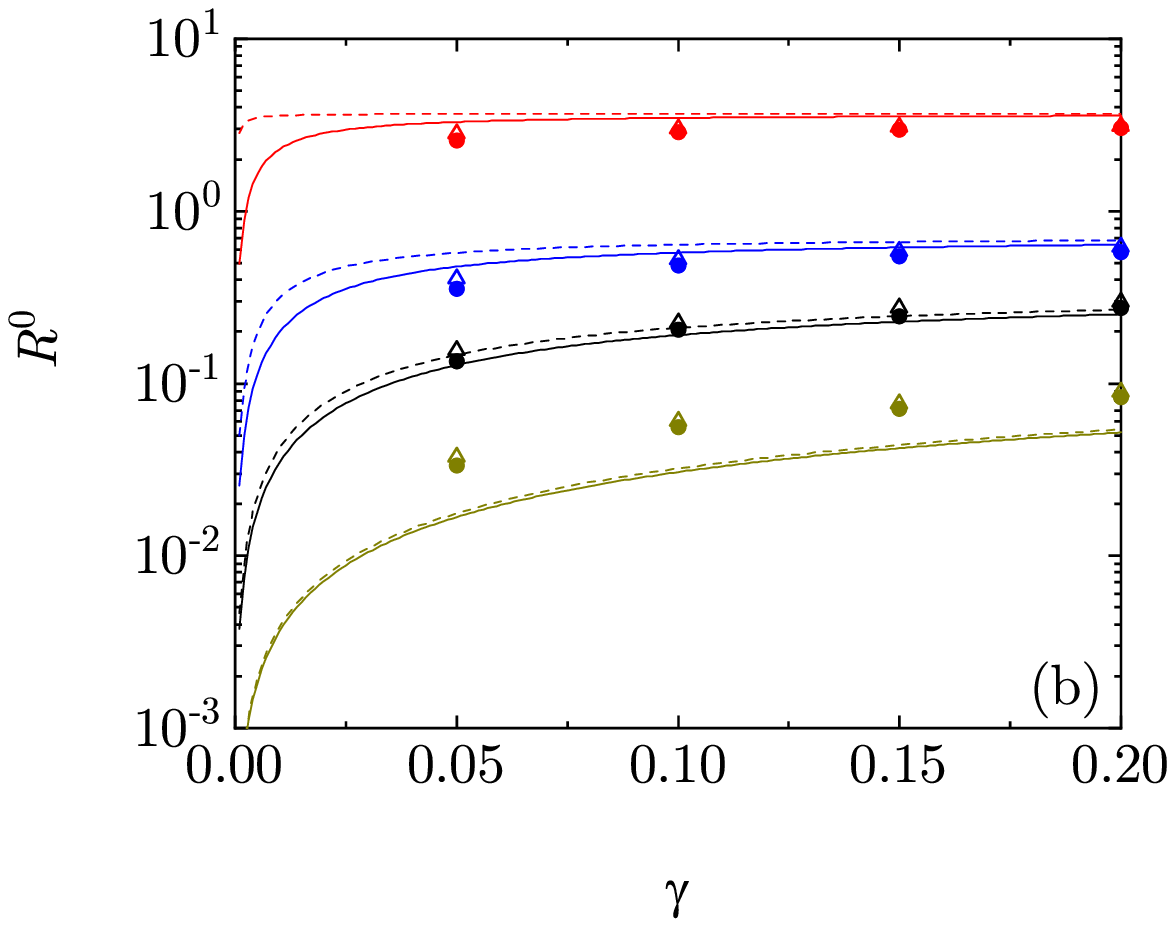}
\caption{\label{fig:phase_diag} Phase diagram in the plane
  $R^{0}\equiv \Delta\theta^0/\Delta a_2^0$ vs $\gamma$ for $d=3$. The
  values of $\zeta_0^*$ are (a) $\zeta_0^*=0.01$ (solid curves and
  closed circles) and $0.1$ (dashed curves and open triangles), and
  (b) $\zeta_0^*=1$ (solid curves and closed circles) and $2$ (dashed
  curves and open triangles). From bottom to top in each panel,
  $\theta_r=0.5$, $1.05$, $2$, and $10$.  In each case, the Mpemba
  effect is present (absent) below (above) the corresponding
  curve. Symbols are obtained by numerically solving Eqs.\
  \eqref{25b-30} with $a_{2A}^0=0.5$ and $a_{2B}^0=-0.35$, whereas
  lines correspond to the analytical prediction given by Eq.\
  \eqref{Delta-max}.}
\end{figure}
A phase diagram in the $(\gamma,R^{0})$ plane can be constructed, as
illustrated in Fig.~\ref{fig:phase_diag}. The line $R^{0}=R^{0}_{\thr}$
separates the regions in which the Mpemba effect is present
($R^{0}<R^{0}_{\thr}$) and absent ($R^{0}>R^{0}_{\thr}$), where $R^{0}_{\thr}$
is defined in Eq.~\eqref{Delta-max}. The range $0<R^{0}<R^{0}_{\thr}$
for which the Mpemba effect emerges increases with $\gamma$,
$\zeta_0^*$, and $\theta_r$. It must be remarked that $\theta_{r}<1$
corresponds to the inverse Mpemba effect, in which the system relaxes
to equilibrium from below the steady temperature $T_{\st}$.

The threshold values $R^{0}_{\thr}$ obtained from the numerical
solution of Eqs.\ \eqref{25b-30} for $a_{2A}^0=0.5$ and
$a_{2B}^0=-0.35$ are also shown in Fig.\ \ref{fig:phase_diag}. The
agreement with the simplified model is quite good, especially in
panel (a), for which $\zeta_{0}^{*}<1$. In panel (b), for which
$\zeta_{0}^{*}\geq 1$, the linearized theory still gives a
semi-quantitative picture and, notably, successfully captures the weak
influence of both $\gamma$ and $\zeta_0^*$ on $R^{0}_{\thr}$ if
$\zeta_0^*\geq 1$. In any case, the linear model overestimates
(underestimates) $R^{0}_{\thr}$ for $\theta_r=10$ and $2$
($\theta_r=1.05$ and $0.5$), as anticipated from Fig.\ \ref{fig:t_c}.

Interestingly, the maximum ratio \emph{relative} to the reference
temperature, $\theta_r^{-1}R^{0}_{\thr}$, keeps increasing with
increasing $\theta_r$. At fixed $\theta_r$, the upper bound of
$R^{0}_{\thr}$ corresponds to the limit $\gamma\to\infty$, which is
independent of $\zeta_0^*$, namely
\begin{equation}
\label{4.10}
\lim_{\gamma\to\infty}R^{0}_{\thr}
=\frac{2(d+2)\theta_r^2/(d-2+12\theta_r)}
{1+\sqrt{1+16(d+2)\theta_r^2/(d-2+12\theta_r)^2}}.
\end{equation}
Notwithstanding, in our
modeling, we are only retaining the first correction,
quadratic in the velocities, in the drag coefficient
$\zeta(v)$. Therefore, from a physical point of view, $\gamma$ is
expected not to very large; otherwise, higher order terms in the
velocity should be incorporated into the drag coefficient.

\subsection{Magnitude of the Mpemba effect}

When the Mpemba effect is present, the temperature difference
$\Delta\theta$ vanishes at the crossover time $t_{c}^{*}$. Since
$\Delta\theta$ also vanishes in the long time limit as
$t^{*}\to\infty$, there must exist a certain time $t^*_m>t_c^*$ where
$|\Delta\theta(t^*)|$ reaches a local maximum. Therefore, one has that
$|\Delta\theta(t^*)|\leq|\Delta\theta(t^*_m)|$ for any time
$t^*>t_c^*$.

\begin{figure*}
\includegraphics[width=\ancho]{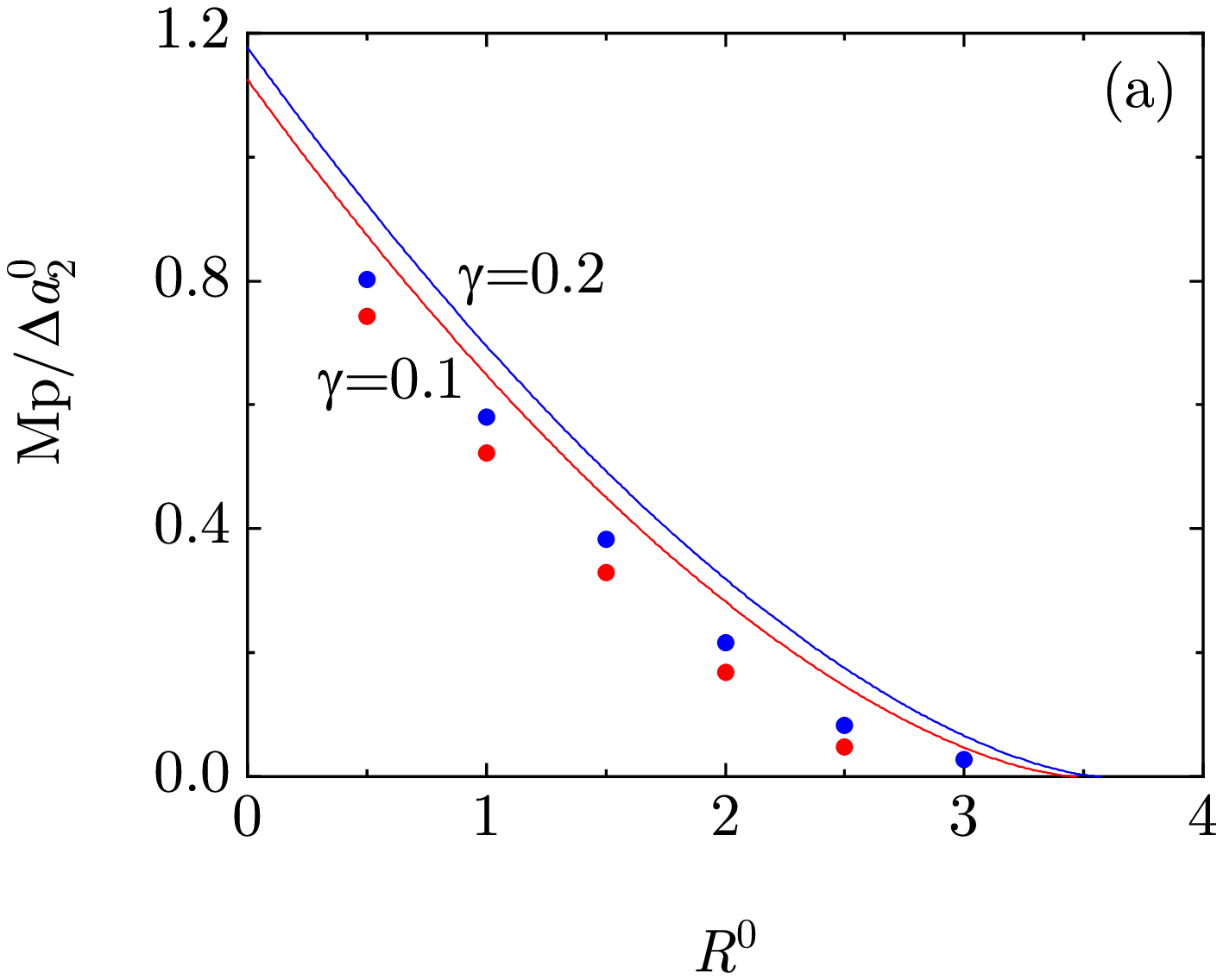}
\includegraphics[width=\ancho]{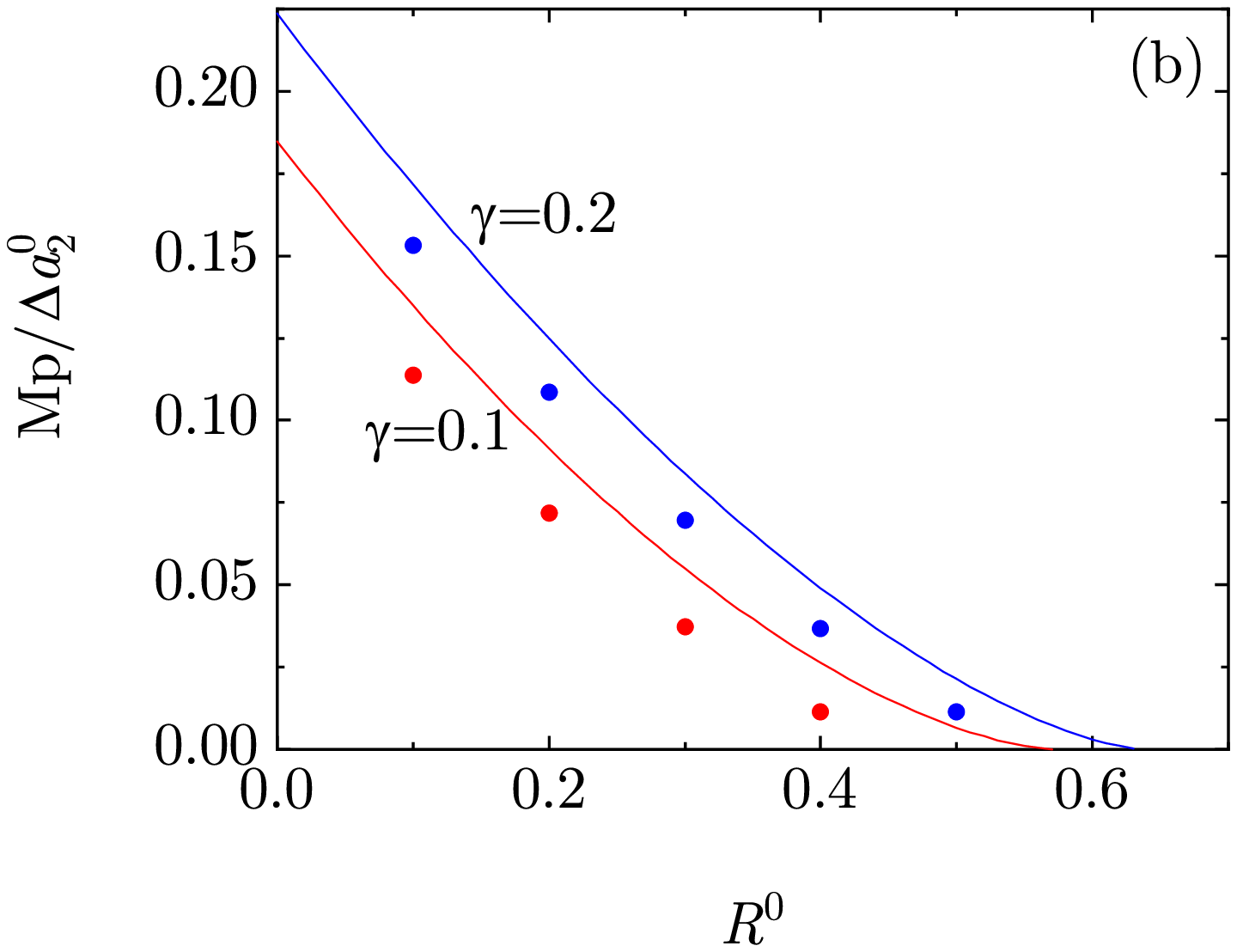}\\
\vspace{2mm}
\includegraphics[width=\ancho]{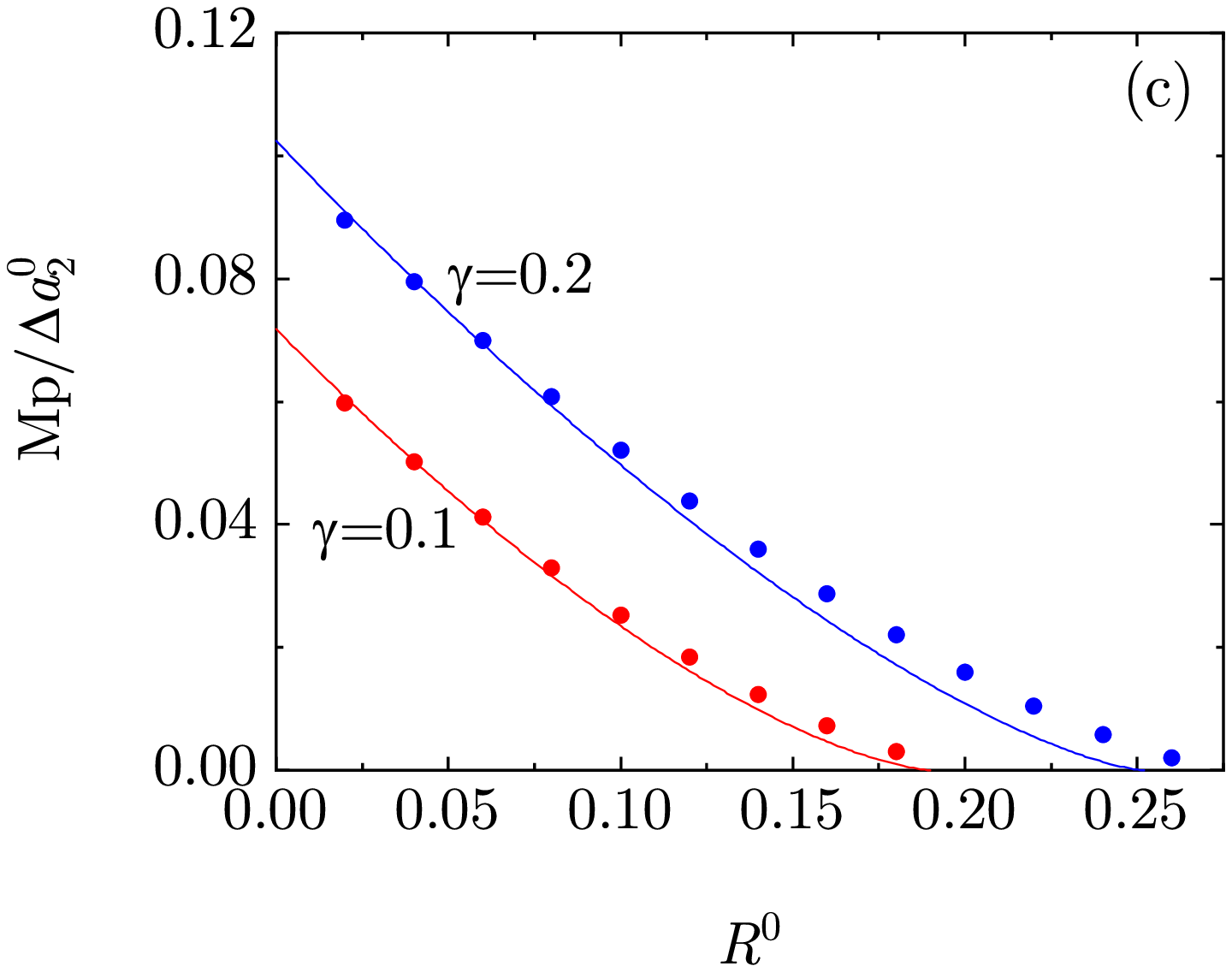}
\includegraphics[width=\ancho]{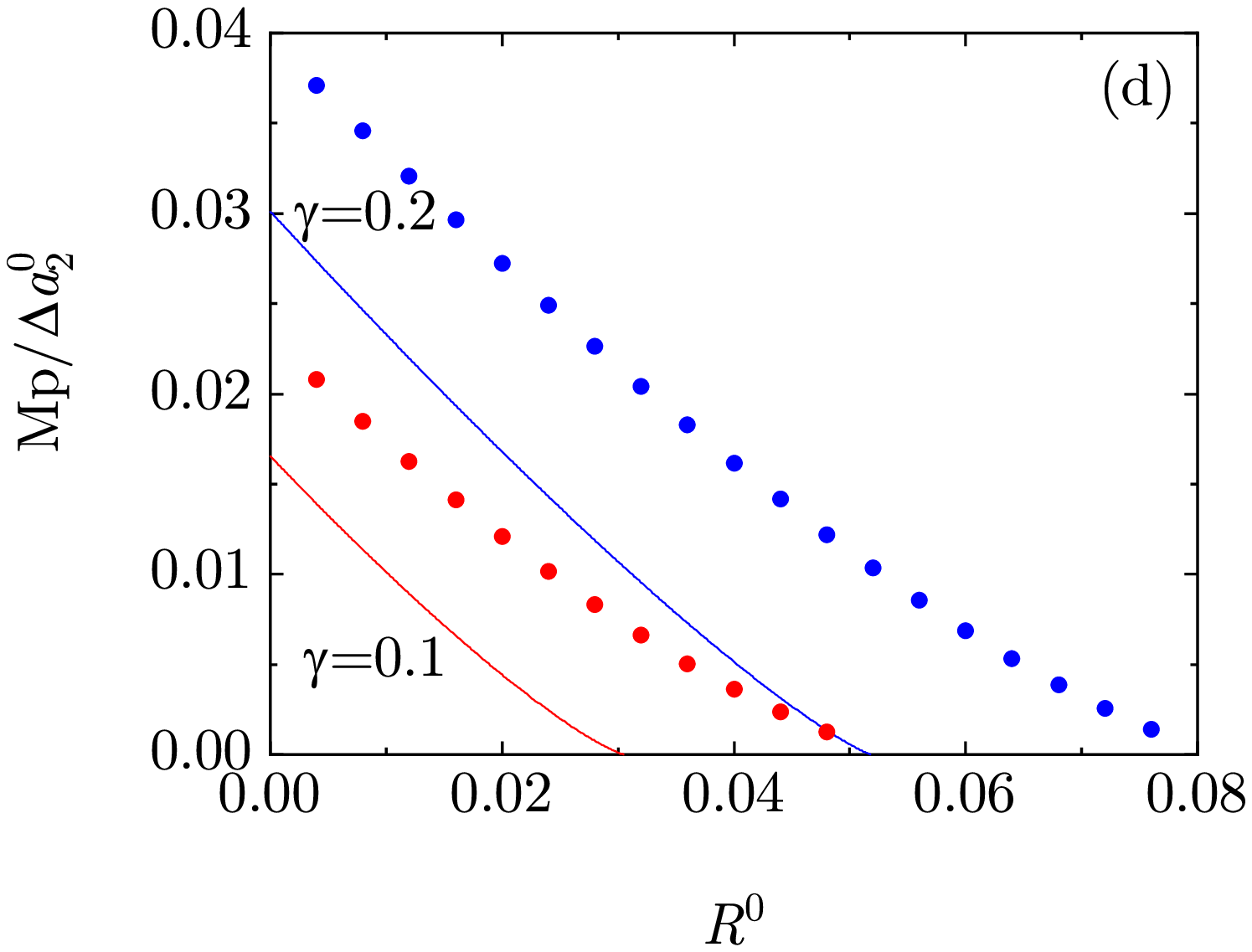}
\caption{\label{fig:Mp} Magnitude of the Mpemba effect,
  $\Mp/\Delta a_2^0$, as a function of
  $R^{0}\equiv \Delta\theta^0/\Delta a_2^0$. The parameter values are
  the same as in Fig.~\ref{fig:t_c}: (i) in all the panels we have
  that $d=3$ and $\zeta_0^*=1$, and (ii) from panel (a) to (d),
  different values of the reference temperature are considered, namely
  $\theta_r=10$, $2$, $1.05$, and $0.5$, with two values of the
  nonlinearity parameter being considered in each panel, $\gamma=0.1$
  (lower curves) and $0.2$ (upper curves).  Again, symbols are
  obtained by numerically solving Eqs.\ \eqref{25b-30} with
  $a_{2A}^0=0.5$ and $a_{2B}^0=-0.35$, whereas lines correspond to the
  analytical prediction given by Eq.\ \eqref{Mp-def}.}
\end{figure*}

The above discussion can be used to introduce a quantitative measure
of the magnitude of the Mpemba effect. Let us define\cite{TLLVPS19}
\begin{equation}
\Mp\equiv |\Delta\theta(t_m^*)|
\end{equation}
as a quantitative measure of its magnitude. From Eqs.\ \eqref{relax-explicit2} one finds
\begin{subequations}
\begin{align}
\label{max-time}
  t_{m}^{*}
  %=& \frac{1}{\delta_\lambda}\ln\left(\frac{\lambda_+}{\lambda_-}\frac{1-\beta  R^0/R^{0}_{\thr}}{1-R^0/R^{0}_{\thr}}\right)\nn
           = &t_c^*+\frac{1}{\delta_\lambda}\ln\frac{\lambda_+}{\lambda_-},\\
  \frac{\Mp}{\Delta
a_2^0}=&\Lambda_{12}\left(\frac{1-R^{0}/R^{0}_{\thr}}{\lambda_{+}}\right)^{\frac{\lambda_{+}}{\delta_\lambda}}\left(\frac{\lambda_{-}}{1-\beta
         R^{0}/R^{0}_{\thr}}\right)^{\frac{\lambda_{-}}{\delta_\lambda}}.
         \label{Mp-def}
\end{align}
\end{subequations}
Thus, the Mpemba magnitude $\Mp$ depends on the initial differences $\Delta\theta^0$ and $\Delta a_2^0$ by  a simple scaling law: the ratio $\Mp/\Delta a_2^0$ is a function of the ratio $R^0\equiv\Delta\theta^0/\Delta a_2^0$. Figure \ref{fig:Mp} shows that the larger the ratio
$R^{0}$ is, the smaller  $\Mp/\Delta a_{2}^{0}$ becomes. Of course,
$\Mp$ vanishes as $R^{0}$ approaches its threshold value
$R^{0}_{\thr}$, as readily seen in Eq.~\eqref{Mp-def}.
Comparison with the values of $\Mp$ obtained from the numerical solution of Eqs.\ \eqref{25b-30} shows that the simple linearized model is qualitatively correct in capturing the dependence of the order of magnitude of $\Mp$  on the parameters of the problem. In agreement with what was observed in Figs.\ \ref{fig:t_c} and \ref{fig:phase_diag}, the linearized model tends to overestimate (underestimate) $\Mp$ for the direct (inverse) Mpemba effect. Figure \ref{fig:Mp}(c) shows that the prediction of $\Mp$ is especially accurate if the initial temperatures are close to the equilibrium one; in that case, $\Mp$ is slightly overestimated for small $R^0$, while it is slightly underestimated as $R^0$ approaches its threshold value $R_\thr^0$.

\subsection{Reliability of the linear theory}

The linear theory we have presented does not apply for all times,
unless the reference temperature $\theta_{r}=\theta_{\st}=1$ and both
initial temperatures $\theta_{A}^{0}$ and $\theta_{B}^{0}$ are close
to the steady state.  Nevertheless, as already stated before, our
linear theory is not the standard linearization around the steady
state but an approximate scheme to obtain a good approximation to the
actual time evolution of the system in the early stage of its
evolution, where the Mpemba effect is expected to come about. This
means that our linear approach does have some limitations, as observed in Figs.\ \ref{fig:t_c}--\ref{fig:Mp}.
According to them, the linearized model becomes more accurate as
$|\theta_r-1|$ and $\zeta_0^*$ decrease.
%, and as $\gamma$ increases. THIS IS NOT SO CLEAR TO ME

While a complete account of the range of validity of the linear
approximation is outside the scope of our paper, quite simple
arguments can be presented in the limit of weak nonlinearity,
$\gamma\to 0^{+}$. The behavior of the threshold
$R^{0}_{\thr}$---below which the Mpemba effect is found--depends on
those of its numerator, $\Lambda_{12}$, and its
denominator, $\lambda_{+}-\Lambda_{11}$. On the one hand, Eq.~\eqref{4.3b} tells us
that the former is linear in $\gamma$, $\Lambda_{12}=O(\gamma)$. On
the other hand, the behavior of  $\lambda_{+}-\Lambda_{11}$ for small $\gamma$ depends
on the sign of the function
\begin{equation}
\label{varphi}
\varphi(\theta_r,\zeta_0^*)\equiv \frac{4(d-1)}{d(d+2)\zeta_0^*}\theta_r^{3/2}-\theta_r+2.
\end{equation}
Specifically, it can be readily shown that
\begin{equation} \lambda_{+}-\Lambda_{11}\approx\left\{
    \begin{array}{ll} \displaystyle
\frac{2\zeta_0^*\varphi(\theta_r,\zeta_0^*)}{\theta_{r}}, &
\varphi(\theta_r,\zeta_0^*)>0, \\ \\ \displaystyle
-\frac{8\zeta_0^*(d+2)\theta_{r}^{3}}{\varphi(\theta_r,\zeta_0^*)}\gamma^{2}, &
\varphi(\theta_r,\zeta_0^*)<0,
    \end{array} \right.
\end{equation}
and thus
\begin{equation}
  \label{eq:R0-estimate}
  R^{0}_{\thr}\sim\left\{
    \begin{array}{ll} \displaystyle (d+2)\theta_{r}^{3}\frac{\gamma}{\varphi(\theta_r,\zeta_0^*)},
& \varphi(\theta_r,\zeta_0^*)>0, \\ \\ \displaystyle
-\frac{1}{4\theta_{r}}\frac{\varphi(\theta_r,\zeta_0^*)}{\gamma}, &
\varphi(\theta_r,\zeta_0^*)<0.

    \end{array} \right.
\end{equation}

\begin{figure}
\includegraphics[width=\ancho]{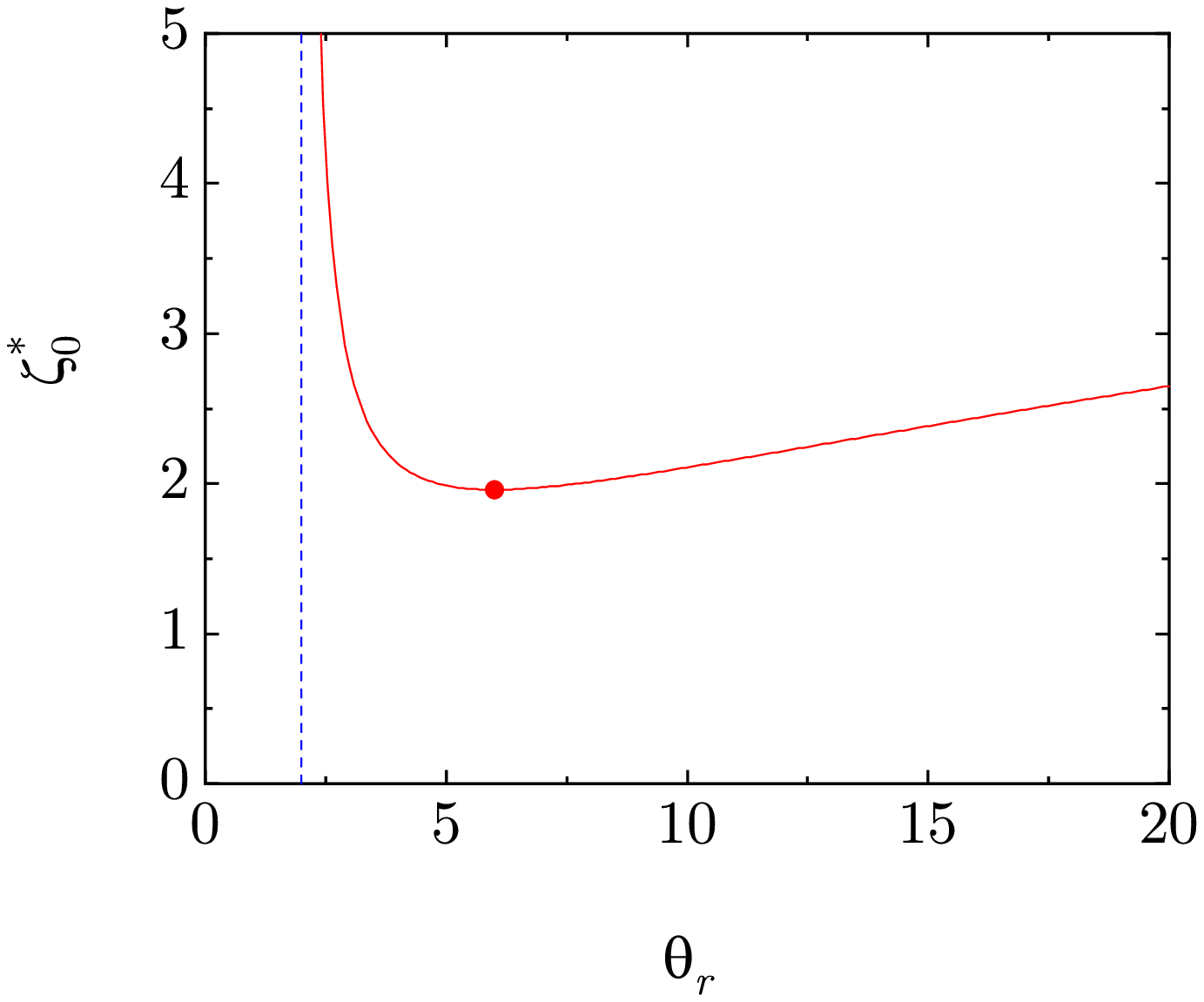}
\caption{\label{fig:locus} Locus $\varphi(\theta_r,\zeta_0^*)=0$ in
  the plane $\zeta_0^*$ vs $\theta_r$ for $d=3$. The
  function $\varphi(\theta_r,\zeta_0^*)$, given by Eq.~\eqref{varphi},
  is negative above (positive below) the curve. The circle represents
  the minimum $\theta_r=6$,
  $\zeta_0^*=\frac{4}{5}\sqrt{6}\simeq 1.96$, whereas the dashed line
  at $\theta_r=2$ is the vertical asymptote of the left branch of the
  locus. Thus, $\zeta_{0}^{*}$ must be large enough, specifically
  larger than its value at the minimum, to allow for negative values
  of $\varphi$. }
\end{figure}

In the limit as $\gamma\to 0^{+}$, the drag becomes linear, the
temperature obeys a closed equation and no Mpemba effect can be
present in the system. This is consistent with the behavior found for
the threshold $R^{0}_{\thr}$ when $\varphi(\theta_r,\zeta_0^*)>0$;
therein, $R^{0}_{\thr}\to 0$. However, $R^{0}_{\thr}\to \infty$
when $\varphi(\theta_r,\zeta_0^*)<0$; this is an unphysical result
that makes us conclude that the simplified linear model
\eqref{relax-explicit2} ceases to be reliable if
$\varphi(\theta_r,\zeta_0^*)<0$ and $\gamma\ll 1$. The locus
$\varphi(\theta_r,\zeta_0^*)=0$ is plotted in Fig.~\ref{fig:locus}.

The above discussion should not be employed to disregard the linear
model directly when $\varphi(\theta_r,\zeta_0^*)<0$; the linearization
can be useful unless $\gamma$ is very small. We can estimate the
value of $\gamma$ for which the linear theory is no longer accurate, by asking the
estimate for
$R^{0}_{\thr}$ in \eqref{eq:R0-estimate} to be large. This leads to
the condition
\begin{equation}
  \label{eq:gamma-l}
  \gamma\ll\gamma_{\ell}\equiv
  \frac{\left|\varphi(\theta_r,\zeta_0^*)\right|}{4\theta_{r}}
  .
\end{equation}

\begin{figure}
\includegraphics[width=\ancho]{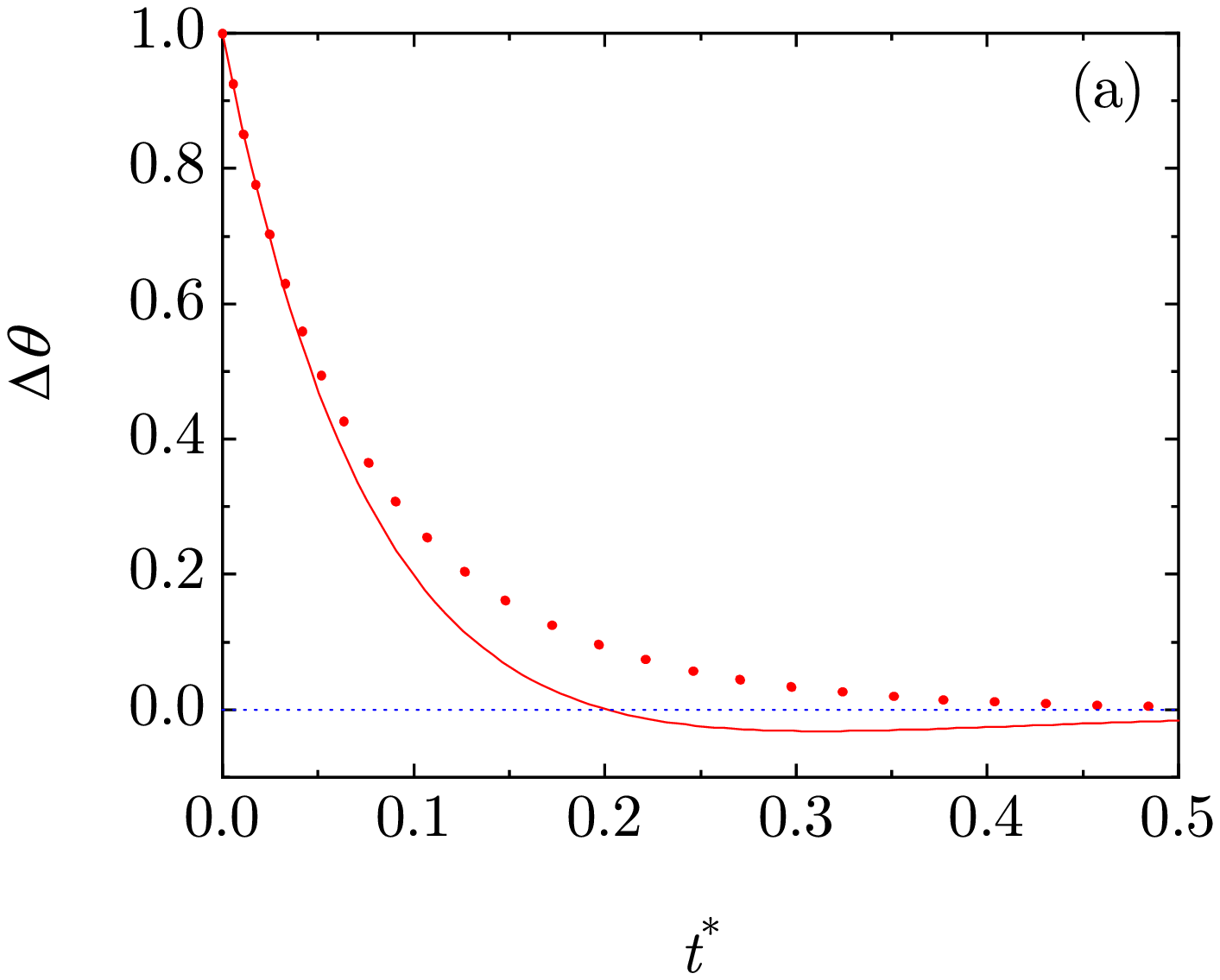}\\
\vspace{2mm}
\includegraphics[width=\ancho]{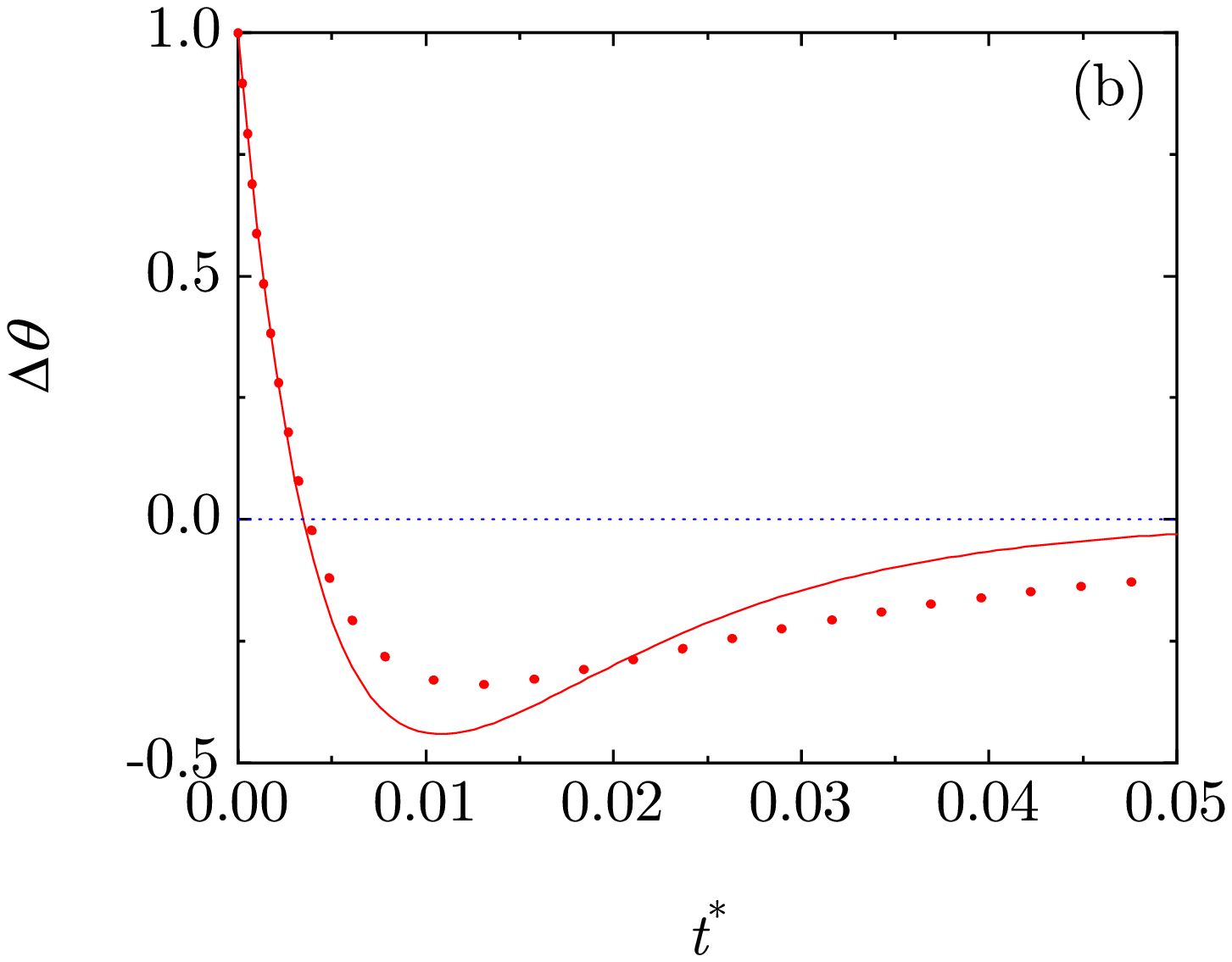}
\caption{\label{fig:Delta} Evolution of
  $\Delta\theta(t^*)=\theta_A(t^*)-\theta_B(t^*)$ for $d=3$,
  $\zeta_0^*=5$. The two panels correspond to small values of
  $\gamma$, namely (a) $\gamma=0.001$ and (b) $\gamma=0.1$. The
  initial states are $\{\theta_A^0,a_{2A}^0\}=\{10,0.5\}$,
  $\{\theta_B^0,a_{2B}^0\}=\{9,-0.35\}$. Circles correspond
  to the numerical solutions of Eqs.\ \eqref{25b-30}, whereas  solid
  lines correspond to the linearized model \eqref{relax-explicit2}
  with $\theta_r=\theta_{B}^0$. The value of $\gamma$ in panel (a)
  verifies the condition \eqref{eq:gamma-l} that controls the failure
  of the linearized theory. }
\end{figure}

We illustrate the above result in Fig.\ \ref{fig:Delta} for the
three-dimensional case, specifically for $\zeta_0^*=5$ and
$\theta_r=9$, in which case we have that
$\varphi(\theta_r,\zeta_0^*)\simeq -4.12$ and
$\gamma_{\ell}\simeq 0.114$. While the (weak) Mpemba effect predicted
by the linearized model with $\gamma=0.001$ is actually absent, the
model succeeds in locating the crossover time if $\gamma=0.1$, which
is quite close to $\gamma_{\ell}$. Note that $\gamma=0.1$ corresponds
to the case in which the mass of the Brownian particles and that of
the surrounding fluid are identical, as shown in Appendix \ref{app-B}.

\section{Conclusions}\label{sec-concl}

We have neatly observed the Mpemba effect in a molecular gas with
nonlinear drag.  For the Mpemba effect---and also for the inverse
Mpemba effect, in which the initially cooler system heats sooner---to
emerge the initially hotter sample must have a sufficiently larger
value of the excess kurtosis $a_{2}^{0}$; the larger $a_{2}^{0}$ is, the
larger the cooling rate becomes. This behavior is completely analogous
to that found in a granular gas of smooth hard spheres.\cite{LVPS17}

The above analysis entails that the Mpemba effect is absent if both
samples, A and B, are initially at equilibrium. In that case,
$a_{2A}^0=a_{2B}^0=0$ and the parameter $R^0$ defined in Eq.\
\eqref{crossover-time} diverges to infinity. For the Mpemba effect to
emerge, we need to prepare the samples in nonequilibrium states before
coupling them to the common reservoir at temperature $T_{\st}$. This
can be achieved, for instance, by temporarily coupling the samples to their
respective reservoirs at temperatures different from $T_\st$.

Analytical predictions have been obtained, in a wide range
of values of the system parameters, within a linearized model. The
linearization is carried around a reference
temperature---specifically, the initial temperature of the sample that
is closer to the equilibrium value, not around the steady temperature.
Therefore, our analytical framework is not limited to near-equilibrium
situations. Within this scheme, we have found semi-quantitatively accurate expressions for
(i) the crossover time, (ii) the maximum value of the initial temperature
difference, and (iii) the magnitude of the Mpemba effect. Also, we
have looked into the limitations of the linearized model, especially
for small values of the parameter $\gamma$ characterizing the
nonlinearity.

% It is worthwhile noting that the Mpemba effect reported here is absent if both systems are initially at equilibrium ($a_{2A}^0=a_{2B}^0=0$)
% with different temperatures ($T_A^0\neq T_B^0$), since in that case
% However, even if starting from an equilibrium state ($a_2^0=0$) with $T^0\neq T_\st$, one has $a_2(t)\neq 0$ (and thus the system is out of equilibrium) during the relaxation stage. Therefore, the Mpemba
% effect can be observed if, for example, both systems start from
% equilibrium but are temporarily coupled to respective reservoirs at
% temperatures  different from $T_\st$, before being coupled to a
% common reservoir at temperature $T_\st$.

This work also opens avenues for further research. It is interesting
to consider in more detail some specific limits of the present model,
which are physically relevant: (i) small nonlinearity $\gamma\ll 1$,
which appears naturally as the first correction to the usual linear
drag, and (ii) time scale separation between viscous drag and
collisions, i.e., either $\zeta_{0}^{*}\ll 1$ or $\zeta_{0}^{*}\gg
1$. In both cases, a systematic---mainly perturbative---analytical
approach seems to be feasible. Also, it is important to deepen our
understanding of aging phenomena in this molecular gas. Specifically,
looking into the Kovacs effect,\cite{K63,KAHR79} which has attracted a
lot of attention
lately,\cite{PB10,BL10,DH11,CWYHM13,PT14,TP14,RP14,BCNO16,KSI17,LGAR17,DS18,LVPS19,LZLDL19,MLTVL20}
is compelling.

Finally, we plan to check the theoretical predictions of this work against computer simulations obtained from the Langevin equation with an interpretation of the multiplicative noise\cite{K94,SSMKD82,vK07,MM12}  consistent with the Enskog--Fokker--Planck equation \eqref{1}.

\acknowledgments
Professor Jason Reese was a brilliant scientist in
the field of kinetic theory, with very important contributions at both
the theoretical and the computational level. With this paper, we
respectfully pay tribute to his memory. A.S. and A.P. acknowledge
financial support from the Spanish Agencia Estatal de Investigaci\'on
through Grants No.\ FIS2016-76359-P and and No.\ PGC2018-093998-B-I00,
respectively. A.S. is also grateful to the Junta de Extremadura
(Spain) for Grant No.\ GR18079. All the grants are partially financed
by the European Regional Development Fund.

\appendix

\section{Hard interaction between Brownian particles and background
  fluid}\label{app-B}

In Refs.~\onlinecite{F07,F14,HKLMSLW17}, the authors consider the
emergence of nonlinear Brownian motion when an ensemble of ``heavy''
Brownian particles (mass $m$, number density $n$) moves in a bath
modeled as a  dilute gas of ``light'' particles (mass $\mbf$,
number density $\nbf$), which is at equilibrium at temperature
$T_{\st}$. A velocity-dependent drag coefficient $\zeta(v)$ is
obtained as an expansion in powers of the mass ratio $\mbf/m$, which is formally
assumed to be small.

The coefficients of the expansion of $\zeta(v)$ are given in terms of
integrals that involve the differential cross section for the
interaction between the Brownian particles and the bath particles.
Explicit expressions for $\zeta(v)$ can be derived when simple
potentials are employed for this interaction.  For example, all
particles are considered to be three-dimensional hard spheres in
Ref.~\onlinecite{HKLMSLW17}, both the Brownian particles and the
particles in the background dilute gas. With this assumption, it is
found that
\begin{equation}
  \label{eq:b1}
  \zeta(v)=\frac{8}{15} \nbf S \sqrt{\frac{2\mbf}{\pi
    k_{B}T_{\st}}}\frac{\mbf v^{2}/2+5k_{B}T_{\st}}{m+\mbf}.
\end{equation}
Above, $S$ is the total cross section, i.e.,
\begin{equation}
  \label{eq:b2}
  S=\frac{\pi\left(\sigma+\sbf\right)^{2}}{4},
\end{equation}
where $\sigma$ and $\sbf$ are the diameters of the Brownian
particles and the background fluid particles, respectively. Equation \eqref{eq:b1} is valid up to order
$(\mbf/m)^{3/2}$ and for not too large velocities, i.e., velocities that
lie in the thermal range $mv^{2}/2k_{B}T_{\st}=O(1)$.

By comparing
Eq.~\eqref{eq:b2} with Eq.~\eqref{eq:zeta-v}, we have
\begin{equation}
  \label{eq:b4}
  \zeta_{0}=\frac{2}{3}\nbf\left(\sigma+\sbf\right)^{2}\sqrt{\frac{\pi
     k_{B}T_{\st}}{m}}\frac{\sqrt{2m\mbf}}{m+\mbf},
\end{equation}
which is proportional to $\sqrt{T_{\st}}$, and
\begin{equation}
  \label{eq:b5}
  \gamma=\frac{\mbf}{10m}.
\end{equation}
which is expected to be small.

In the framework developed in this paper, we measure time in terms of
the number of collisions of Brownian particles among
themselves. Therefore, our evolution equations involve the
dimensionless low-velocity drag coefficient
$\zeta_{0}^{*}=\zeta_{0}\tau_{\st}$, introduced in
Eq.~\eqref{eq:z0*}. For hard spheres ($d=3$) in the Boltzmann limit
[$g(\sigma)=1$], the characteristic time $\tau_{\st}$
for Brownian--Brownian collisions is
\begin{equation}
  \label{eq:b6}
  \tau_{\st}^{-1}=2n\sigma^{2}\sqrt{\frac{\pi k_{B}T_{\st}}{m}}.
\end{equation}
Straightforward algebraic manipulation leads to
\begin{equation}
  \label{eq:b7}
  \zeta_{0}^{*}=\frac{2\nbf}{3n}\left(1+\frac{\sbf}{\sigma}\right)^{2
  }\frac{\sqrt{5\gamma}}{1+10\gamma}.
\end{equation}
Therefore, we have that $\zeta_{0}^{*}$ depends on three dimensionless
quantities---the density ratio $\nbf/n$, the diameter ratio
$\sbf/\sigma$, and the mass ratio $\mbf/m$, as measured by
$\gamma$. This means that, even  in the ``natural'' heavy Brownian limit
$\mbf/m\ll 1$ or $\gamma\ll 1$, $\zeta_{0}^{*}$ varies across a large
range of values. For a given problem, its specific value depends on
$\nbf/n$ and $\sbf/\sigma$, but not on the steady temperature; the
ratio of time scales associated with the viscous drag and
Brownian--Brownian collisions is thus independent of the temperature of the
bath.

The simple expressions for $\gamma$ and $\zeta_{0}^{*}$,
Eqs.~\eqref{eq:b5} and \eqref{eq:b7}, hold for hard-sphere interaction
in the Boltzmann limit. More complicated behaviors may be found in
other situations, but we expect the qualitative picture derived here
to be still valid. As a consequence, while the range of $\gamma$ is somehow limited,
that of $\zeta_{0}^{*}$ is not necessarily so. This explains why we
have restricted ourselves to $\gamma\leq 0.2$ throughout the paper but
treated $\zeta_{0}^{*}$ as an independent parameter, which may attain
both small and large values.

\section{Solution of the linearized system}\label{app-A}

Our starting point is the nonlinear system~\eqref{25b-30}, written in
the Sonine approximation. First, we introduce the deviation of the
temperature from a certain reference temperature $\theta_{r}$ by
\begin{equation}
  \label{eq:A1}
  \Psi(t^{*})=\theta(t^{*})-\theta_{r}.
\end{equation}
Second, we linearize Eq.~\eqref{25b-30} with respect to $\Psi(t^{*})$
and $a_{2}(t^{*})$. Note that Eqs.~\eqref{25b-30} are linear in $a_{2}$ but
nonlinear in $\theta$ and, in addition, the ``coefficients'' of
$a_{2}$ are functions of $\theta$. The result is
\begin{equation}
\label{linmod}
\begin{pmatrix} \dot{\Psi}\\
     \dot{a}_{2}
\end{pmatrix}=-
\begin{pmatrix}
  \Lambda_{11}&\Lambda_{12}\\
  \Lambda_{21}&\Lambda_{22}
\end{pmatrix}\cdot
\begin{pmatrix} \Psi\\
a_{2}
\end{pmatrix}
+\begin{pmatrix}
  C_1\\
  C_2
\end{pmatrix},
\end{equation}
where
\begin{subequations}
\label{As}
\bal
C_1=&2\zeta_0^*\left(1-\theta_r\right)\left[1+\gamma(d+2)\theta_r\right],\\
C_2=&8\zeta_0^*\gamma \left(1-\theta_r\right),
\eal
\end{subequations}
and  $\Lambda_{ij}$ has been defined in Eq.~\eqref{4.3}. The solution of the simplified linear model \eqref{linmod} yields
\begin{subequations}
\label{relax-explicit}
\begin{align}
\label{temp-relax-explicit}
\Psi(t^*)=&D_{1}+\frac{(\lambda_{+}-\Lambda_{11})\left(\Psi^{0}-D_{1}\right)-
\Lambda_{12} \left(a_{2}^{0}-D_{2}\right)}{\delta_\lambda e^{\lambda_{-}t^*}}\nn
  &
-\frac{(\lambda_{-}-\Lambda_{11})\left(\Psi^{0}-D_1\right)-
\Lambda_{12} \left(a_{2}^{0}-D_2\right)}{\delta_\lambda e^{\lambda_{+}t^*}},\\
\label{a2-relax-explicit}
a_2(t^*)=&D_{2}+\frac{(\lambda_{+}-\Lambda_{22})\left(a_{2}^{0}-D_2\right)-
\Lambda_{21} \left(\theta^{0}-D_1\right)}{\delta_\lambda e^{\lambda_{-}t^*}}\nn
  &
-\frac{(\lambda_{-}-\Lambda_{22})\left(a_{2}^{0}-D_2\right)-
\Lambda_{21} \left(\theta^{0}-D_1\right)}{\delta_\lambda e^{\lambda_{+}t^*}},
\end{align}
\end{subequations}
where $\lambda_{\pm}$ are the eigenvalues of the matrix $\mathsf{\Lambda}$
defined in Eq.~\eqref{4.5}  and we have defined the parameters
\begin{equation}
\label{Bs}
D_1=\frac{\Lambda_{22}C_1-\Lambda_{12}C_2}{\Lambda_{11}\Lambda_{22}-\Lambda_{12}\Lambda_{21}},\quad
D_2=\frac{\Lambda_{11}C_2-\Lambda_{21}C_1}{\Lambda_{11}\Lambda_{22}-\Lambda_{12}\Lambda_{21}}.
\end{equation}
Neither $\Psi$ nor $a_{2}$ reaches its actual equilibrium value  in the long-time limit, unless $\theta_{r}=1$. This is not a
problem for the analysis carried out in the main text, because this
linear theory is only used for the early stage of the time evolution,
in which the Mpemba effect may emerge.

Now, let us consider two different initial states: A, with initial
values of the temperature and the excess kurtosis
$\{\theta_{A}^{0},a_{2A}^{0}\}$, and B, with initial values
$\{\theta_{B}^{0},a_{2B}^{0}\}$. The linear theory makes it possible
to write analytical predictions for the differences between their
respective time evolutions, i.e.,
$\Delta\theta(t^{*})=\theta_{A}(t^{*})-\theta_{B}(t^{*})=\Psi_{A}(t^{*})-\Psi_{B}(t^{*})$
and $\Delta a_{2}(t^{*})=a_{2A}(t^{*})-a_{2B}(t^{*})$. Making use of
Eq.~\eqref{relax-explicit}, we arrive precisely at
Eq.~\eqref{relax-explicit2} in the main text.

\section*{DATA AVAILABILITY}

The data that support the findings of this study are available from the corresponding author upon reasonable request.

\section*{REFERENCES}

\bibliography{Granular}

\end{document}